\documentclass[oldversion]{aa}
\usepackage{epsfig}
\usepackage{txfonts}
\begin{document}
   \title{Galaxy-cluster gas-density distributions of the
   Representative {\it XMM-Newton} Cluster Structure Survey (REXCESS)}

   \author{J.H. Croston
          \inst{1}
          \and
          G.W. Pratt\inst{2}
	  \and
	  H. B\"ohringer\inst{2}
	  \and
	  M. Arnaud\inst{3}
	  \and
	  E. Pointecouteau\inst{4}
	  \and
	  T. J. Ponman\inst{5}
	  \and
	  A. J. R. Sanderson\inst{5}
	  \and
	  R. F. Temple\inst{5}
	  \and
	  R. G. Bower\inst{6}
	  \and
	  M. Donahue\inst{7}
}

   \offprints{J.H. Croston}

   \institute{Centre for Astrophysics Research, Science and Technology
              Research Institute, University of Hertfordshire,
              Hatfield, AL10 9AB, UK
              \email{J.H.Croston@herts.ac.uk}
	      \and
	      MPE Garching, Giessenbachstrasse, 85748 Garching, Germany
              \and
	      Laboratoire AIM, DAPNIA/Service d'Astrophysique - CEA/DSM - CNRS - Universit\'{e} Paris Diderot, B\^{a}t. 709, CEA-Saclay, F-91191 Gif-sur-Yvette Cedex, France
	      \and
	      CESR,  9 av. du colonel Roche, BP4346, 31028 Toulouse Cedex 4, France
	      \and
	      School of Physics and Astronomy, University of
              Birmingham, Edgbaston, Birmingham, B15 2TT, UK
	      \and
	      Institute for Computational Cosmology, University of Durham, Durham, DH1 3LE, UK
              \and
	      Physics and Astronomy Department, Michigan State University, East
              Lansing, MI 48824-2320, USA
	      \\
             }

   \date{}

   \abstract{ We present a study of the structural and scaling
     properties of the gas distributions in the intracluster medium
     (ICM) of 31 nearby ($z < 0.2$) clusters observed with {\it
     XMM-Newton}, which together comprise the Representative {\it
     XMM-Newton} Cluster Structure Survey (REXCESS). In contrast to
     previous studies, this sample is unbiased with respect to X-ray
     surface brightness and cluster dynamical state, and it fully
     samples the cluster X-ray luminosity function. The clusters cover
     a temperature range of 2.0 -- 8.5 keV and possess a variety of
     morphologies. The sampling strategy allows us to compare clusters
     with a wide range of central cooling times on an equal footing.
     We applied a recently developed technique for the deprojection
     and PSF-deconvolution of X-ray surface brightness profiles to
     obtain non-parametric gas-density profiles out to distances
     ranging between $0.8R_{500}$ and $1.5 R_{500}$. We scaled the gas
     density distributions to allow for the systems' differing masses
     and redshifts. The central gas densities differ greatly from
     system to system, with no clear correlation with system
     temperature. At intermediate radii ($\sim 0.3 R_{500}$), the
     scaled density profiles show much less scatter, with a clear
     dependence on system temperature. We find that the density at
     this radius scales proportionally to the square root of
     temperature, consistent with the presence of an entropy excess as
     suggested in previous literature. However, at larger scaled radii
     this dependence becomes weaker: clusters with $kT > 3$ keV scale
     self-similarly, with no temperature dependence of gas-density
     normalisation. The REXCESS sample allows us to investigate the
     correlations between cluster properties and dynamical state. We
     find no evidence of correlations between cluster dynamical state
     and either the gas density slope in the inner regions or
     temperature, but do find some evidence of a correlation between
     dynamical state and outer gas density slope. We also find a weak
     correlation between dynamical state and both central gas normalisation and inner cooling
     times, but this is only significant at the 10\% level. We
     conclude that, for the X-ray cluster population as a whole, both
     the central gas properties and the angle-averaged, large-scale
     gas properties are linked to the cluster dynamical state. We also
     investigate the central cooling times of the clusters. While the
     cooling times span a wide range, we find no evidence of a
     significant bimodality in the distributions of central density,
     density gradient, or cooling time. Finally, we present the gas
     mass-temperature relation for the REXCESS sample, finding that
     $h(z)M_{gas} \propto T^{1.99\pm0.11}$, which is consistent with
     the expectation of self-similar scaling modified by the presence
     of an entropy excess in the inner regions of the cluster and
     consistent with earlier work on relaxed cluster samples. We
     measure a logarithmic intrinsic scatter in this relation of $\sim
     10\%$, which should be a good measure of the intrinsic scatter in
     the $M_{gas}$--$T$ relation for the cluster population as a
     whole.}

\authorrunning{J.H. Croston et al.}
\titlerunning{REXCESS gas density distributions}

   \maketitle
%

\section{Introduction}

The X-ray emitting gas in galaxy clusters contains the signatures of
important evolutionary processes such as mergers, AGN activity, galaxy
interactions and tidal stripping. The impact of gas physical processes
on the observable X-ray properties of galaxy clusters must be fully
understood in order to use galaxy clusters to test the predictions of
structure formation models, and to understand the relationship between
AGN activity, galaxy evolution and the evolution of large-scale
structure. The global scaling relations between cluster observables
such as X-ray luminosity/temperature and cluster mass must be well
constrained in the local Universe so that cluster evolution to high
redshifts can be investigated; this is not possible without a good
understanding of the processes that lead to the observed scatter in
these relationships.

Since the first evidence that the gas and dark matter content of
galaxy clusters does not scale according to the simplest self-similar
predictions (e.g. Edge \& Stewart 1991; Arnaud \& Evrard 1999),
non-gravitational heating processes have been thought to play an
important role in determining the X-ray properties of galaxy clusters.
Recent results have shown that the dark matter component of galaxy
clusters does scale self-similarly, in good agreement with theoretical
predictions (e.g. Pointecouteau et al. 2005; Vikhlinin et al. 2006).
It has also been shown that the departure of the observed
$L_{X}/T_{X}$ relation from theoretical predictions is due to
increased entropy in cluster gas (e.g. Ponman et al. 2003; Pratt,
Arnaud \& Pointecouteau 2006). A combination of radiative cooling and
feedback processes associated with galaxy formation, either from
galaxy winds or AGN activity associated with central supermassive
black holes, is the most widely accepted explanation for these results
(e.g. Voit 2005). Cluster mergers are also expected to have an
important effect on their observed X-ray properties: for example,
simulations have shown that cluster merger processes are likely to
have an important effect on the $L_{X}/T_{X}$ relation (Rowley et al.
2004), although cooling and heating processes in the inner regions
may be the dominant source of scatter (O'Hara et al. 2006).

Radial surface brightness profiles and gas density profiles have been
the key means of obtaining information about the structure and scaling
properties of the intracluster medium (ICM) since the advent of X-ray
imaging of clusters, providing an essential tool to enable study of
three-dimensional gas distributions, and, together with temperature,
the study of the gas entropy distributions and the total mass profiles
via the assumption of hydrostatic equilibrium. {\it ROSAT} studies
(e.g. Neumann \& Arnaud 1999) suggested that a $\beta$ model was an
adequate parametrization of the X-ray surface brightness distribution
outside the core region in galaxy clusters, with clear evidence for
self-similarity, but a large dispersion in the central regions. The
core regions of clusters are now much better resolved with {\it
XMM-Newton} and {\it Chandra}, and more complex models are required to
fit their surface brightness profiles (e.g. Pratt \& Arnaud 2002;
Pointecouteau et al. 2004; Vikhlinin et al. 2006). There is also some
evidence that the slope of the gas density profile may steepen at
large radii in some clusters (e.g. Viklinin et al. 1999; Neumann 2005;
Vikhlinin et al. 2006). Constraints on the radial distribution of hot
gas in the galaxy cluster population are crucial both for accurate
estimation of total cluster masses and for studies of gas entropy.
Spherically symmetric methods for constraining the gas distributions
of galaxy clusters have obvious limitations when applied to samples
that are not selected for regularity; however, as a straightforward
method that is easy to apply to both observations and simulations,
they remain an essential tool for studying cluster properties in three
dimensions.

\begin{table*}[t]
\begin{minipage}[t]{17.5cm}
\caption{Cluster properties. Columns: (1) Cluster ID, (2) redshift,
  (3) column density in $10^{20}$ cm$^{-2}$, (4) X-ray temperature
  from a {\it mekal} fit in the $[0.15 - 0.75] R_{500}$ region in keV,
  (5) X-ray temperature from a {\it mekal} fit in the
  $[0.15-1]R_{500}$ region in keV, (6) abundance obtained from a
  {\it mekal} fit in the $[0.15 - 1] R_{500}$ region, in units of
  solar abundance, (7) $R_{500}$ in kpc, iteratively determined from
  the $M_{500}-Y_{X}$ relation of Arnaud, Pointecouteau \& Pratt
  (2007) as described in the text, (8) logarithm of gas mass to
  $R_{500}$ determined as described in the text in $M_{\sun}$, (9)
  inner slope of the gas density profile determined in the region with
  $r < 0.05 R_{500}$, (10) outer slope of the gas density profile
  determined in the region$[0.3 - 0.8]R_{500}$ as defined in the
  text.}
\label{sample}
\centering
\renewcommand{\footnoterule}{}  
\begin{tabular}{llcccccccc}
\hline \hline
Cluster& z & $N_{H}$& $T_{X} (0.75R_{500})$&$T_{X} (R_{500})$ & $Z$ & $R_{500}$ & log($M_{gas}$) & $\alpha$ ($<0.05R_{500}$) & $\beta$ ($[0.3-0.8]R_{500}$)\\
\hline
RXCJ0003.8$+$0203 & 0.0924 & 3.0 & 3.87$_{-0.10}^{+0.10}$ & 3.64$_{-0.09}^{+0.09}$ & 0.27$_{-0.04}^{+0.04}$ & 876.69 & 13.298$\pm$0.006 & 0.55$\pm$0.00 & 0.63$\pm$0.01 \\
RXCJ0006.0$-$3443 & 0.1147 & 1.1 & 5.18$_{-0.20}^{+0.20}$ & 4.60$_{-0.16}^{+0.21}$ & 0.34$_{-0.06}^{+0.06}$ & 1059.31 & 13.642$\pm$0.010 & 0.48$\pm$0.00 & 0.50$\pm$0.01 \\
RXCJ0020.7$-$2542 & 0.1410 & 2.1 & 5.55$_{-0.13}^{+0.13}$ & 5.24$_{-0.15}^{+0.15}$ & 0.18$_{-0.04}^{+0.04}$ & 1045.30 & 13.606$\pm$0.008 & 0.21$\pm$0.00 & 0.74$\pm$0.01 \\
RXCJ0049.4$-$2931 & 0.1084 & 1.9 & 3.03$_{-0.12}^{+0.12}$ & 2.79$_{-0.11}^{+0.11}$ & 0.26$_{-0.04}^{+0.05}$ & 807.79 & 13.225$\pm$0.009 & 0.48$\pm$0.01 & 0.65$\pm$0.01 \\
RXCJ0145.0$-$5300 & 0.1168 & 2.8 & 5.63$_{-0.14}^{+0.14}$ & 5.51$_{-0.16}^{+0.16}$ & 0.30$_{-0.05}^{+0.05}$ & 1089.28 & 13.674$\pm$0.009 & 0.17$\pm$0.00 & 0.56$\pm$0.00 \\
RXCJ0211.4$-$4017 & 0.1008 & 1.6 & 2.07$_{-0.05}^{+0.05}$ & 2.02$_{-0.06}^{+0.06}$ & 0.27$_{-0.02}^{+0.02}$ & 685.04 & 12.991$\pm$0.008 & 0.70$\pm$0.02 & 0.61$\pm$0.03 \\
RXCJ0225.1$-$2928 & 0.0604 & 1.6 & 2.67$_{-0.13}^{+0.13}$ & 2.61$_{-0.16}^{+0.16}$ & 0.69$_{-0.09}^{+0.11}$ & 693.91 & 12.874$\pm$0.012 & 0.04$\pm$0.01 & 0.56$\pm$0.00 \\
RXCJ0345.7$-$4112 & 0.0603 & 1.8 & 2.30$_{-0.06}^{+0.09}$ & 2.15$_{-0.08}^{+0.08}$ & 0.37$_{-0.04}^{+0.05}$ & 688.40 & 12.919$\pm$0.013 & 1.19$\pm$0.05 & 0.62$\pm$0.02 \\
RXCJ0547.6$-$3152 & 0.1483 & 2.1 & 6.06$_{-0.14}^{+0.14}$ & 5.68$_{-0.11}^{+0.11}$ & 0.27$_{-0.03}^{+0.03}$ & 1133.74 & 13.768$\pm$0.005 & 0.26$\pm$0.01 & 0.62$\pm$0.01 \\
RXCJ0605.4$-$3518 & 0.1392 & 4.5 & 4.91$_{-0.11}^{+0.11}$ & 4.81$_{-0.12}^{+0.12}$ & 0.31$_{-0.04}^{+0.04}$ & 1045.94 & 13.659$\pm$0.008 & 0.87$\pm$0.04 & 0.70$\pm$0.01 \\
RXCJ0616.8$-$4748 & 0.1164 & 5.1 & 4.17$_{-0.11}^{+0.11}$ & 4.16$_{-0.12}^{+0.12}$ & 0.31$_{-0.04}^{+0.04}$ & 939.16 & 13.452$\pm$0.008 & 0.61$\pm$0.04 & 0.50$\pm$0.02 \\
RXCJ0645.4$-$5413 & 0.1644 & 6.5 & 7.27$_{-0.18}^{+0.18}$ & 6.97$_{-0.19}^{+0.19}$ & 0.22$_{-0.04}^{+0.04}$ & 1279.98 & 13.994$\pm$0.007 & 0.49$\pm$0.01 & 0.60$\pm$0.01 \\
RXCJ0821.8$+$0112 & 0.0822 & 4.2 & 2.84$_{-0.10}^{+0.10}$ & 2.44$_{-0.12}^{+0.12}$ & 0.28$_{-0.04}^{+0.05}$ & 755.86 & 13.071$\pm$0.010 & 0.68$\pm$0.02 & 0.64$\pm$0.02 \\
RXCJ0958.3$-$1103 & 0.1669 & 5.4 & 6.30$_{-0.44}^{+0.50}$ & 5.85$_{-0.40}^{+0.45}$ & 0.25$_{-0.00}^{+0.00}$ & 1077.39 & 13.648$\pm$0.016 & 0.64$\pm$0.02 & 0.81$\pm$0.01 \\
RXCJ1044.5$-$0704 & 0.1342 & 3.6 & 3.57$_{-0.05}^{+0.05}$ & 3.52$_{-0.05}^{+0.05}$ & 0.26$_{-0.02}^{+0.02}$ & 931.85 & 13.518$\pm$0.008 & 1.11$\pm$0.06 & 0.69$\pm$0.01 \\
RXCJ1141.4$-$1216 & 0.1195 & 3.2 & 3.54$_{-0.05}^{+0.05}$ & 3.40$_{-0.06}^{+0.06}$ & 0.38$_{-0.03}^{+0.03}$ & 885.24 & 13.385$\pm$0.012 & 0.98$\pm$0.05 & 0.62$\pm$0.01 \\
RXCJ1236.7$-$3354 & 0.0796 & 5.5 & 2.73$_{-0.01}^{+0.09}$ & 2.57$_{-0.03}^{+0.11}$ & 0.42$_{-0.04}^{+0.04}$ & 753.50 & 13.078$\pm$0.009 & 0.41$\pm$0.01 & 0.61$\pm$0.01 \\
RXCJ1302.8$-$0230 & 0.0847 & 1.7 & 3.44$_{-0.07}^{+0.07}$ & 2.92$_{-0.07}^{+0.09}$ & 0.26$_{-0.03}^{+0.03}$ & 842.12 & 13.247$\pm$0.013 & 1.04$\pm$0.06 & 0.50$\pm$0.02 \\
RXCJ1311.4$-$0120 & 0.1832 & 1.8 & 8.44$_{-0.12}^{+0.12}$ & 8.24$_{-0.13}^{+0.13}$ & 0.26$_{-0.02}^{+0.02}$ & 1319.18 & 14.019$\pm$0.004 & 0.56$\pm$0.01 & 0.71$\pm$0.01 \\
RXCJ1516.3$+$0005 & 0.1181 & 4.7 & 4.48$_{-0.07}^{+0.07}$ & 4.18$_{-0.08}^{+0.08}$ & 0.25$_{-0.03}^{+0.03}$ & 989.86 & 13.548$\pm$0.008 & 0.44$\pm$0.00 & 0.65$\pm$0.01 \\
RXCJ1516.5$-$0056 & 0.1198 & 5.4 & 3.74$_{-0.09}^{+0.10}$ & 3.40$_{-0.08}^{+0.08}$ & 0.30$_{-0.03}^{+0.03}$ & 927.02 & 13.472$\pm$0.009 & 0.51$\pm$0.01 & 0.41$\pm$0.01 \\
RXCJ2014.8$-$2430 & 0.1538 & 13.1 & 5.73$_{-0.10}^{+0.10}$ & 5.63$_{-0.11}^{+0.11}$ & 0.27$_{-0.03}^{+0.03}$ & 1155.29 & 13.843$\pm$0.013 & 0.88$\pm$0.15 & 0.64$\pm$0.01 \\
RXCJ2023.0$-$2056 & 0.0564 & 5.3 & 2.72$_{-0.09}^{+0.09}$ & 2.46$_{-0.12}^{+0.12}$ & 0.20$_{-0.04}^{+0.04}$ & 739.51 & 13.014$\pm$0.010 & 0.46$\pm$0.00 & 0.58$\pm$0.01 \\
RXCJ2048.1$-$1750 & 0.1475 & 4.7 & 5.01$_{-0.11}^{+0.11}$ & 4.59$_{-0.08}^{+0.08}$ & 0.22$_{-0.03}^{+0.03}$ & 1077.96 & 13.730$\pm$0.008 & 0.01$\pm$0.00 & 0.51$\pm$0.01 \\
RXCJ2129.8$-$5048 & 0.0796 & 2.2 & 3.88$_{-0.14}^{+0.14}$ & 3.64$_{-0.12}^{+0.16}$ & 0.46$_{-0.06}^{+0.06}$ & 900.60 & 13.350$\pm$0.008 & 0.37$\pm$0.00 & 0.46$\pm$0.01 \\
RXCJ2149.1$-$3041 & 0.1184 & 2.3 & 3.50$_{-0.07}^{+0.07}$ & 3.40$_{-0.08}^{+0.08}$ & 0.26$_{-0.03}^{+0.03}$ & 886.56 & 13.393$\pm$0.010 & 0.75$\pm$0.06 & 0.56$\pm$0.01 \\
RXCJ2157.4$-$0747 & 0.0579 & 3.6 & 2.76$_{-0.07}^{+0.07}$ & 2.30$_{-0.06}^{+0.10}$ & 0.28$_{-0.03}^{+0.03}$ & 751.45 & 13.047$\pm$0.008 & 0.42$\pm$0.00 & 0.37$\pm$0.01 \\
RXCJ2217.7$-$3543 & 0.1486 & 1.1 & 4.65$_{-0.08}^{+0.10}$ & 4.45$_{-0.09}^{+0.09}$ & 0.21$_{-0.03}^{+0.03}$ & 1022.61 & 13.638$\pm$0.007 & 0.45$\pm$0.00 & 0.60$\pm$0.01 \\
RXCJ2218.6$-$3853 & 0.1411 & 1.4 & 6.16$_{-0.19}^{+0.19}$ & 5.88$_{-0.15}^{+0.20}$ & 0.34$_{-0.05}^{+0.05}$ & 1130.13 & 13.747$\pm$0.006 & 0.31$\pm$0.00 & 0.66$\pm$0.01 \\
RXCJ2234.5$-$3744 & 0.1510 & 1.2 & 7.30$_{-0.12}^{+0.12}$ & 6.95$_{-0.14}^{+0.14}$ & 0.23$_{-0.03}^{+0.03}$ & 1283.21 & 13.984$\pm$0.007 & 0.13$\pm$0.01 & 0.71$\pm$0.01 \\
RXCJ2319.6$-$7313 & 0.0984 & 2.9 & 2.52$_{-0.07}^{+0.07}$ & 2.48$_{-0.08}^{+0.08}$ & 0.31$_{-0.04}^{+0.04}$ & 788.73 & 13.238$\pm$0.010 & 0.81$\pm$0.07 & 0.51$\pm$0.02 \\
\hline
\hline
\end{tabular}
\end{minipage}
\end{table*}

In this paper we examine the gas distributions in 31 clusters drawn
from the Representative {\it XMM-Newton} Cluster Structure Survey
(REXCESS). Full details of the sample selection function can be found
in B\"ohringer et al. (2007). To achieve a statistically well-defined
sample which fully samples the X-ray luminosity function and is
unbiased with respect to dynamical state, a sample of 33 clusters was
constructed from the REFLEX catalogue (B\"ohringer et al. 2004) based
on the following criteria: redshift $z < 0.2$; close to homogeneous
coverage of luminosity space; $kT > 2.0$ keV; detectable with {\it
XMM-Newton} to a radius of $\sim R_{500}$; and distances selected to
optimise the extent of the cluster within the {\it XMM-Newton}
field-of-view. In addition, a firm detection of more than 30 photons
in the original RASS detection, and a low column density towards the
source were also required. Observations of the full sample of 31
clusters plus 2 archive observations have now been completed. The
analysis presented here consists of the 31 clusters for which it is
reasonable to carry out a 1-dimensional analysis. We exclude the
clusters RXCJ2152$-$1942 (Abell 2384B) and RXCJ0956$-$1004 (Abell
901/902) which are highly irregular/multiple cluster systems and hence
cannot be analysed in this way.
 
Throughout this paper we adopt a $\Lambda$CDM cosmology with $H_{0} =
70$ km s$^{-1}$ Mpc$^{-1}$, $\Omega_{M} = 0.3$ and $\Omega_{\Lambda} =
0.7$.

\section{Data analysis}

\subsection{Data preparation}

Table~\ref{sample} lists the global properties of the cluster sample
and the details of the {\it XMM-Newton} observations. Observations
were retrieved and reprocessed with the {\it XMM-Newton} Science
Analysis System (SAS) version 7.0, ensuring up to date calibration
across the sample. Data sets were cleaned for flares, PATTERN-selected
and corrected for vignetting as described in Pratt et al. (2007). The
background used for the present analysis consists of custom event
files generated from data taken in Filter Wheel Closed (FWC) mode,
which correspond to an accurate representation of the particle and
instrumental background present in {\it XMM-Newton}
observations. These background data sets were cleaned,
PATTERN-selected and vignetting corrected as above, then recast to
have the same aspect as the observation data files. The background
event lists were rescaled to the source quiescent count rate in the
[10-12] and [12-14] keV range for EMOS and EPN cameras, respectively,
by adjusting the WEIGHT column in each background event file.

Sources other than the target object were detected in a broad band
([0.3-10.0] keV) coadded EPIC image using the SAS wavelet detection
task {\tt ewavdetect}, with a detection threshold set at $5
\sigma$. After visual screening, detected sources were excluded from
the event file for all subsequent analysis.

\subsection{Surface brightness and gas density profiles}
\label{profiles}
Vignetting-corrected surface brightness profiles were extracted for
each camera in the 0.3 - 2.0 keV band from source and scaled
background event files in $3\farcs3$ bins out to a radius of
15\arcmin. The profiles were centred on the peak in X-ray surface
brightness. The background surface brightness profiles were then
subtracted, and the background subtracted profiles from the three
cameras co-added. No weighting for camera response was applied at this
stage. We then applied the second stage of background subtraction to
account for the X-ray background. As the REXCESS sample was chosen
with specific field-of-view criteria, all of the observations include
an outer region which can be used to measure accurately the cosmic
X-ray background (CXB) component. The residual background level due to
the CXB in the co-added, background subtracted surface brightness
profile was determined by a careful visual analysis of the region
where the profile flattens due to background domination; the mean
level in this region was then calculated and subtracted, and the
resulting profile rebinned to a significance of $3\sigma$ per bin.

Deprojected, PSF-corrected emission measure profiles were obtained
from the surface brightness profiles using the non-parametric method
described in Croston et al. (2006). The response matrices used were
obtained using the Ghizzardi et al. (2001) parametrisation of the {\it
XMM-Newton} PSF, weighted by the contribution of each camera to the
combined profile. The profiles were then converted to gas density by
calculating a global conversion factor for each profile in XSPEC using
the global temperatures listed in Table~\ref{sample}, which are
spectroscopic temperatures estimated in the $[0.15 < R < 1]\,R_{500}$
aperture. A correction factor to take into account radial variations
of temperature and abundance was calculated for each radial bin using
a parameterisation of the projected temperature and abundance
profiles, as detailed in Pratt \& Arnaud (2003)\footnote{The use of
projected values here is acceptable, as the temperature dependence of
emissivity in the 0.3 - 2.0 keV band is $< 5 \%$, and there are no
significant differences between projected and deprojected abundance.}.
The surface brightness profiles and corresponding gas density profiles
for each cluster are presented in the Appendix. The gas density
profiles in tabular form are included in the electronic version of
this paper.

\section{Results}

\subsection{Global properties of cluster gas density profiles}

Fig.~\ref{nprofs} shows the 31 cluster gas density profiles superposed
to allow an investigation of their global properties. The individual
surface brightness profiles (in the energy range 0.3 - 2.0 keV) and
gas density profiles are given in the Appendix. There is a
considerable amount of scatter in the unscaled profiles (top left
panel), which is unsurprising. Structure formation models predict that
the mean cluster density scales with $\rho_{c} (z)$ and so according
to $h(z)^{2}$ and predict a universal profile shape when the radial
coordinate is scaled according to the cluster mass. We use a scaling
radius of $R_{500}$, defined as the radius enclosing a mean
overdensity of 500 times the critical density. We therefore expect
that gas density profiles scaled by $h(z)^{-2}$, and in radial units
of $r/R_{500}$ should coincide.

The physical distances were converted to scaled radius using the
values of $R_{500}$ given in Table~\ref{sample}, which were determined
iteratively as described by Kravtsov et al. (2006), based on the
$M_{500}-Y_{X}$ relation of Arnaud, Pointecouteau \& Pratt (2007) (see
also Maughan et al. 2007). Here $Y_{X}$ is the product of the gas mass
within $R_{500}$ and the temperature in the $[0.15-1]R_{500}$ region.
This approach allows us to determine $R_{500}$ directly from the data
without first fitting total mass profiles. For self-consistency, we
prefer to use a scaling relation determined from {\it XMM-Newton}
results, rather than relying on results determined from lensing
masses, which have large individual uncertainties and are only
applicable to the high mass regime, or from simulations that do not
yet reproduce all of the observed properties of the cluster
population. A comparison between the empirical relation of Arnaud et
al. (2007) and the simulated relation of Nagai et al. (2007) shows
only a small normalisation offset ($< 8 \%$), so that the only
difference to our results if the simulated relation were used would be
to shift all the profiles in the direction of smaller radii. It would
therefore have no effect on any of our conclusions, which are all
based on the relative profiles.

The total gas mass to a radius of $R_{500}$ was determined for the
sample by integrating over the gas density profiles. For profiles
where the data does not extend to $R_{500}$ (9/31 clusters) we
extrapolated the profiles from the outer bin of the density profile
(never less than $0.8 R_{500}$) assuming a power law slope in log
space, determined by a fit to the outer four bins. The errors on gas
mass were determined by combining in quadrature the errors obtained
from Monte Carlo simulation over the statistical errors in the surface
brightness profiles and the errors due to the temperature
uncertainties.

The scatter in gas density at $0.3R_{500}$ is $\sigma/\langle n_{e}
\rangle = 0.265$. We then scaled the gas densities according to their
expected evolution with redshift: $n_{e} \propto n_{0} h(z)^{2}$
(shown in the middle left panel of Fig.~\ref{nprofs}), which reduced
the scatter at $0.3R_{500}$ to $\sigma/ \langle n_{e} \rangle =
0.236$. The bottom right panel of Fig.~\ref{nprofs} shows the relative
dispersion as a function of radius for this scaling (in red), which
drops from $\sim 100\%$ in the inner regions to a minimum of $\sim
15\%$ at $0.6 R_{500}$, before increasing slightly in the outer
regions.

\begin{figure*}[tp]
\begin{center}
\resizebox{0.9\hsize}{!} {
\epsfig{figure=9154f1a.eps}
\hspace{4mm}
\epsfig{figure=9154f1b.eps}}

\vspace{5mm}
\resizebox{0.9\hsize}{!} {
\epsfig{figure=9154f1c.eps}
\hspace{4mm}
\epsfig{figure=9154f1d.eps}}

\vspace{5mm}
\resizebox{0.9\hsize}{!} {
\epsfig{figure=9154f1e.eps}
\hspace{4mm}
\epsfig{figure=9154f1f.eps}}
\caption{\label{nprofs}Gas density profiles for the entire sample. Top left:
  unscaled profiles; top right: profiles of unscaled density with
  scaled radius; middle left: profiles of density scaled for redshift
  evolution; middle right: profiles of density scaled according to
  $T^{0.525}$ as implied by modified self-similar S-T scaling; bottom
  left: density profiles with a representation of the 1-sigma scatter
  of the sample; bottom right: the relative dispersion
  ($\sigma/\langle n_{e} \rangle$) as a function of radius for the
  cluster sample with profiles scaled according to expected evolution
  with redshift (black) and by $T^{0.525}$ (red).}
  \end{center}

\end{figure*}

\subsection{Dependence of gas density on temperature} 
\label{tdep}

\begin{figure}
\epsfig{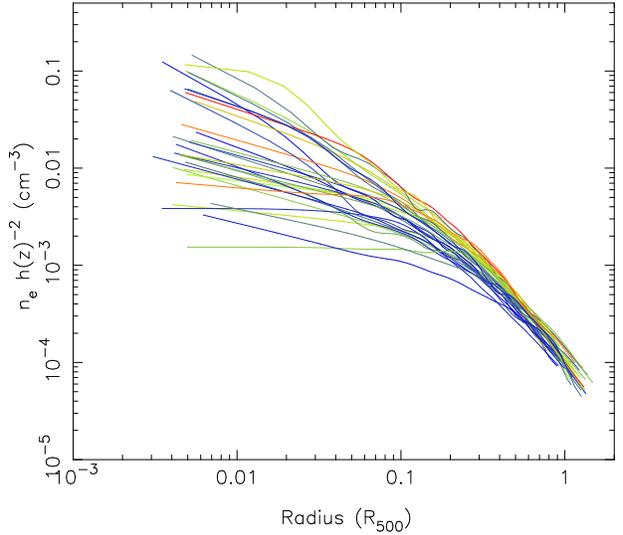}
\caption{Gas density distributed scaled according to expected
  evolution with redshift, and coded by temperature so that blue
  corresponds to a temperature of $\sim 2.0$ keV and red a temperature
  of $\sim 8$ keV.}
\label{colour}
\end{figure}

\begin{figure*}
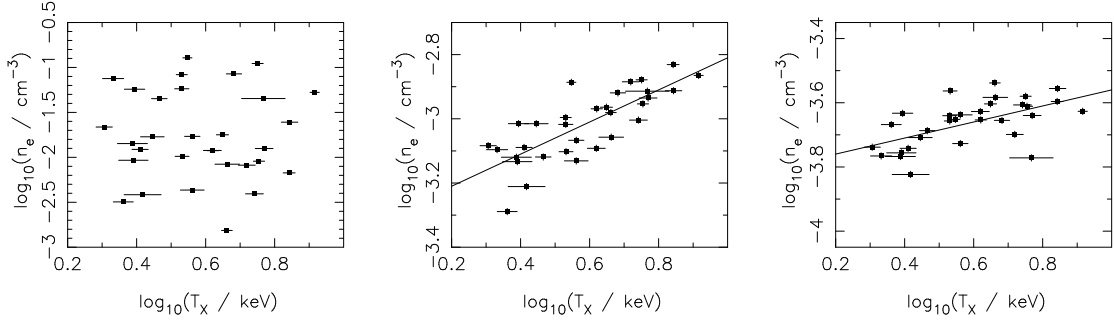

\centering
\resizebox{0.8\hsize}{!} {
\epsfig{figure=9154f3a.eps}
\hspace{2cm}
\epsfig{figure=9154f3b.eps}
\hspace{2cm}
\epsfig{figure=9154f3c.eps}}
\caption{Plots of gas density normalisation vs. cluster temperature at
  $0.03R_{500}$ (left), $0.3R_{500}$ (middle) and at $0.7R_{500}$
  (right), for the redshift-scaled density profiles. For the latter
  two cases, lines of best fit obtained as described in the text are
  overplotted.}
\label{norm}
\end{figure*}

Fig.~\ref{colour} shows the evolution-corrected gas density profiles
colour-coded by cluster temperature. At the smallest radii probed by
this data, the dispersion in profiles is uncorrelated with temperature
(e.g. left hand panel of Fig.~\ref{norm}). We investigate the origin of
this scatter in \S4.3. At intermediate radii ($\sim 0.3 R_{500}$),
there is a clear systematic bias with system temperature, although
this trend becomes weaker again at the maximum radii probed by the
data. Fig.~\ref{norm} also quantifies the relationship between gas
density normalisation at a radius of $0.3 R_{500}$ and at $0.7
R_{500}$. Results are shown for the redshift-scaled gas density
profiles; however, using the unscaled profiles does not significantly
alter the result. There is evidence for a correlation in both cases,
although the trend is weaker at $0.7 R_{500}$. We obtained a null
hypothesis probability $< 0.0001$ on a Spearman rank test for the
comparison at $0.3\,R_{500}$ and a higher null hypothesis probability
of $0.004$ at $0.7\,R_{500}$. We carried out orthogonal linear
regression, finding a slope of $0.50\pm0.08$ for the relationship
between $n_{e} (0.3\,R_{500})$ and $T$.

The departure of the density from the simple scaling is best discussed
in terms of the radial entropy profiles of the systems. To be consistent
with previous work, we define entropy as $S = T / n_{e}^{2/3}$.  
The observed departure from simple scaling of the entropy-temperature
relation for clusters (e.g. Finoguenov et al.\ 2002,  Ponman et al. 2003) 
implies a temperature dependence of the gas density normalisation 
such that $n_{e} \propto T^{0.525}$. This is in excellent agreement 
with the trend we see at $0.3\,R_{500}$. However, at
$0.7\,R_{500}$, the correlation is weaker and we find a flatter
slope of $0.25\pm0.06$. It is apparent from the righthand panel of
Fig.~\ref{norm} that the correlation at $0.7R_{500}$ is mainly due to
lower densities in the systems below $kT \sim 3.0$ keV; at higher
temperatures the relation appears to be flat. This suggests that the
``entropy excess'' may extend to larger scaled radii in cooler
systems, consistent with the expectation that non-gravitational
processes have a greater effect at the low mass end of the cluster
population.

We then scaled the density profiles by $T^{0.525}$ (in addition to the redshift scaling discussed above), as shown in the
middle right panel of Fig.~\ref{nprofs}. As shown in the bottom right
panel, the relative dispersion decreases for this scaling (shown in
red) out to a radius of $\sim 0.5 R_{500}$, reaching a minimum of
$\sigma/\langle n_{e} \rangle = 0.133$ at $0.5 R_{500}$. Beyond $\sim
0.5 R_{500}$ the relative dispersion for this scaling is slightly
higher than for the redshift-evolution-only profiles. This result is
consistent with the weaker correlation between gas density and
temperature at $0.7 R_{500}$, supporting the conclusion that an
entropy excess is likely to be significant in the cluster inner
regions only.

\subsection{Radial dependence of density profile slope}

We investigated the radial dependence of the density profile slope,
defined as the logarithmic gradient of density with radius, and
determined from the deprojected profiles, with a binning chosen to
correspond to a typical temperature profile for the sample. In
Fig.~\ref{slopeplot}, we present the superposed plots of ${\rm d}
log(n_{e}) / {\rm d} log(r)$ for the cluster sample, colour-coded by
temperature in the same way as for Fig.~\ref{colour}. There is a large
scatter at all radii of the profiles, which in many cases are not very
smooth. The scatter at a radius of $0.3R_{500}$ is $\sigma/ \langle
\alpha \rangle = 0.20$. It is evident from Fig.~\ref{slopeplot} that
in most cases $\alpha$ appears still to be increasing in the outermost
regions (e.g. at a distance of $\sim 0.8 - 1.0 R_{500}$) in contrast
to the power-law behaviour expected for a $\beta$-model profile. This
is in agreement with earlier work (e.g. Vikhlinin et al. 1999; Neumann
2005), and has important implications for cluster mass estimates
obtained using analytical models. We have fully taken into account
systematic uncertainties in the background level in determining the
outer slopes: the systematic uncertainty in the background level is
typically $\sim 2 - 3\%$, leading to an uncertainty in outer slope
that is negligible compared to the statistical uncertainties on
surface brightness. In one case, RXCJ2234.5$-$3744, the background
uncertainty is $\sim 14\%$ due to residual flare contamination, which
leads to a $\sim 3 \%$ error in the $[0.3-0.8]R_{500}$ slope. In
Fig~\ref{slopeplot}, the mean slope is shown only out to $0.9R_{500}$,
where the effects of systematic uncertainty in the background level
are negligible. Our conclusions relating to the gas density gradient
are therefore robust.

\begin{figure}
\centering
\epsfig{figure=9154fig4.eps,width=0.9\columnwidth}
\caption{Profiles of $\alpha$, the density profile slope for the
  entire cluster sample, colour-coded by cluster temperature as in
  Fig.~\ref{colour}. Thick red line indicates the mean profile.}
\label{slopeplot}
\end{figure}

\begin{figure*}
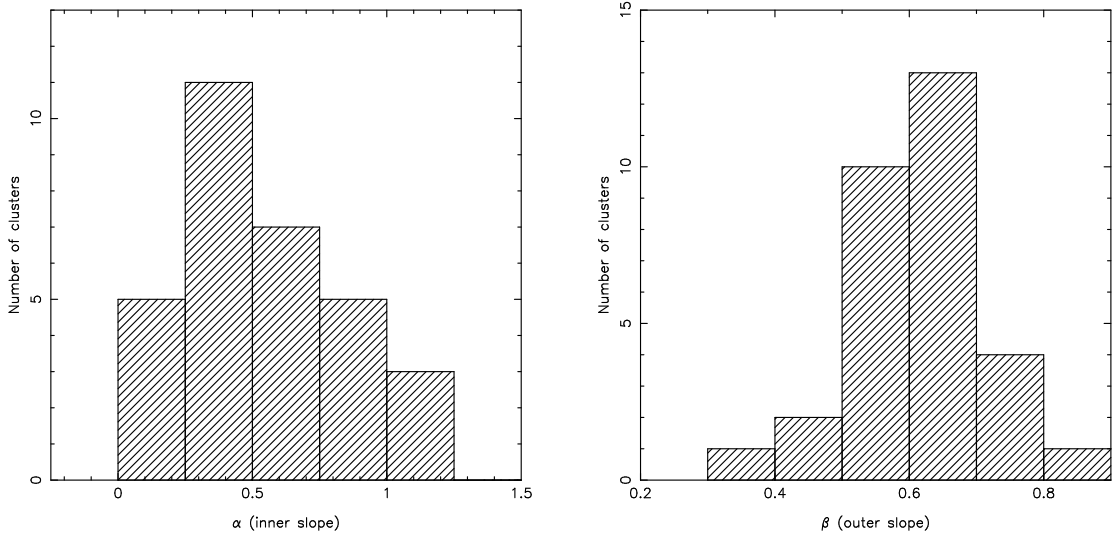

\centering
\resizebox{0.8\hsize}{!} {
\epsfig{figure=9154f5a.eps}
\hspace{2cm}
\epsfig{figure=9154f5b.eps}}
\caption{Histograms of the distribution of inner logarithmic slope
  ($\alpha_{<0.05}$) and outer slope ($\beta_{0.3-0.8}$) for the density profiles scaled
  according to their expected evolution with redshift.}
\label{hist}
\end{figure*}

\subsection{Parametrization of cluster structure}
\label{struc}

\begin{figure}
\centering
\epsfig{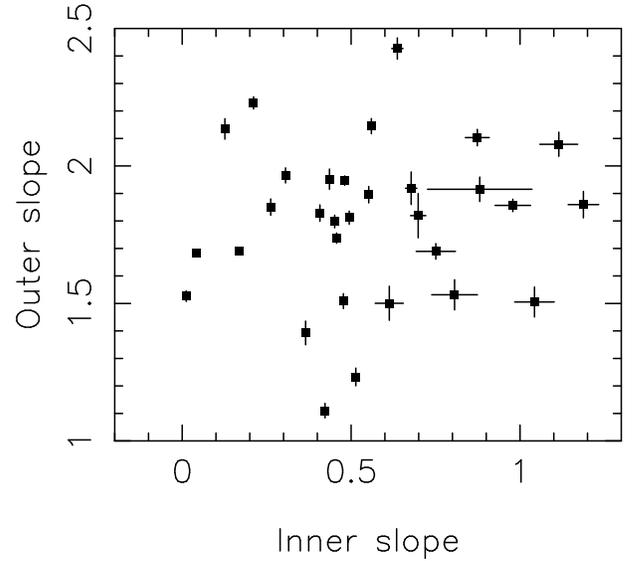}
\caption{Comparison of the inner and outer slopes of the density
  profile ($\alpha_{<0.05}$ and $\alpha_{0.3-0.8}$, respectively) for
  each cluster}
\label{innerouter}
\end{figure}

In order to further investigate the distribution of cluster structure
and the dependence of cluster structure on other properties such as
gas temperature and dynamical state, we first considered the outer
logarithmic slope of the gas density profiles. We chose to use the
region of $[0.3 - 0.8] R_{500}$, since, as shown in Fig.~\ref{nprofs}
and Fig.~\ref{slopeplot}, all of the profiles have steepened from
their inner slopes by $0.3R_{500}$, and all extend to at least
$0.8R_{500}$. The use of a fixed outer radius minimizes the
possibility of bias due to brighter, and hence higher temperature,
clusters extending to larger radii. We express the slope as a $\beta$
value, where $\beta$ is defined as 1/3 times the slope, for comparison
with other works using analytical models. The outer slope in the
region $[0.3 - 0.8] R_{500}$ is hereafter referred to as
$\alpha_{0.3-0.8}$, or $\beta_{0.3-0.8}$ when expressed as a $\beta$
value. We also measured the inner logarithmic slope within
$0.05R_{500}$, before any of the profiles have begun to steepen
significantly, which is hereafter referred to as $\alpha_{<0.05}$.
Both $\alpha_{0.3-0.8}$ and $\alpha_{<0.05}$ were measured by a simple
linear least-squares fit in log space over the relevant region. As
discussed above, the effects of systematic uncertainties in the
background level have been fully taken into account and are less than
the statistical uncertainties in nearly all cases.

In Fig.~\ref{hist} (top) we plot the distribution of $\beta_{0.3-0.8}$
and $\alpha_{<0.05}$ for the gas density profiles scaled according to
expected evolution with redshift. The mean value for $\alpha_{<0.05}$
is $-0.5\pm0.3$, with the large scatter reflecting the range from
clusters with central cores to those that are highly peaked (using a
fixed inner boundary in scaled radius does not reduce the scatter or
significantly alter the mean slope). We are confident that the values
of $\alpha_{0.3-0.8}$ are unaffected by PSF-correction or resolution
issues: in Croston et al. (2006) we showed that the PSF correction is
very accurate beyond $\sim 0.05$ arcmin, and we find that
$\alpha_{0.3-0.8}$ is uncorrelated with redshift or angular scale. The
mean value of $\alpha_{0.3-0.8}$ is $-1.80\pm0.28$ (corresponding to
$\beta_{0.3-0.8} = 0.60\pm0.10$), which is consistent with other
studies, e.g. Neumann \& Arnaud (1999), Ota \& Mitsuda (2004),
Vikhlinin et al. (2006). The scatter in $\alpha_{0.3-0.8}$ is
considerably smaller than for $\alpha_{<0.05}$; however, a few
clusters are exceptionally flat in their outer regions (three clusters
have $\beta < 0.5$). There is no evidence for bimodality in $\alpha_{<0.05}$. We also investigated whether $\alpha_{<0.05}$ and $\alpha_{0.3-0.8}$, the inner and outer slopes,
are correlated. As shown in Fig.~\ref{innerouter}, no such correlation
appears to exist, indicating that in general the dynamical evolution
of the angle-averaged gas distribution in the central and outer
regions of the clusters are not closely connected.

The REXCESS sample includes clusters with evidence for irregularity,
and in some cases the choice of the centre for the profiles is very
dependent on the radius used for centroiding (see
Section~\ref{dynstate}). We therefore examined the effect of this
uncertainty on the structural parameters of the density profiles by
considering three of the most irregular clusters, RXCJ2048$-$1750,
RXCJ2129$-$5048 and RXCJ1516$-$0056. For each of these clusters,
surface brightness profiles were extracted using centroid positions
obtained within radii of $0.3, 0.5$ and $1.0$ times $R_{500}$, which
were deprojected in the same way as for the original profiles. We
found that the choice of centroid unsurprisingly has an important
effect on the measured value of $\alpha_{<0.05}$, introducing a
scatter of between 40 -- 67 percent; however, the results for the
inner two choices of centroiding region, more appropriate for studying
the slope in the inner regions, were in reasonable agreement. The
choice of centroid does not appear to introduce significant
uncertainty in $\alpha_{0.3-0.8}$: the dispersion between the
three cases ranged from 2 -- 6 percent in these ``worst-case''
clusters. We comment below on the effect of centroid choice on other
results.

\subsection{Dependence of cluster structure on temperature }
\label{struc}

\begin{figure*}
\centering
\resizebox{0.8\hsize}{!} {
\epsfig{figure=9154f7a.eps}
\hspace{2cm}
\epsfig{figure=9154f7b.eps}}
\caption{Plots of the inner (left) and outer (right) slopes
  ($\alpha_{<0.05}$ and $\beta_{0.3-0.8}$), respectively, versus the
  global cluster temperature, $T_{X}$.}
\label{slopeT}
\end{figure*}

In Fig.~\ref{slopeT} we plot the inner and outer density slopes
  ($\alpha_{<0.05}$ and $\beta_{0.3-0.8}$) against the cluster
  temperature in the region $[0.15 - 1] R_{500}$. $\beta_{0.3-0.8}$
  appears to correlate with global temperature. Based on a Spearman
  rank test, we find a $\sim 2$ percent probability that such a
  correlation could occur by chance. There is no significant correlation
  between $\alpha_{<0.05}$ and the global cluster temperature.
  It is clear from Fig.~\ref{slopeplot} that the slope does not depend
  on temperature over most of the radial range. It is only at radii
  beyond $\sim 0.3 R_{500}$ that the slopes of the cooler profiles are
  preferentially below the mean, with hotter profiles mainly above the
  mean.

\section{Discussion}

\subsection{Total gas mass and the $M_{gas}$--$T$ relation}

\begin{figure}
\centering
\epsfig{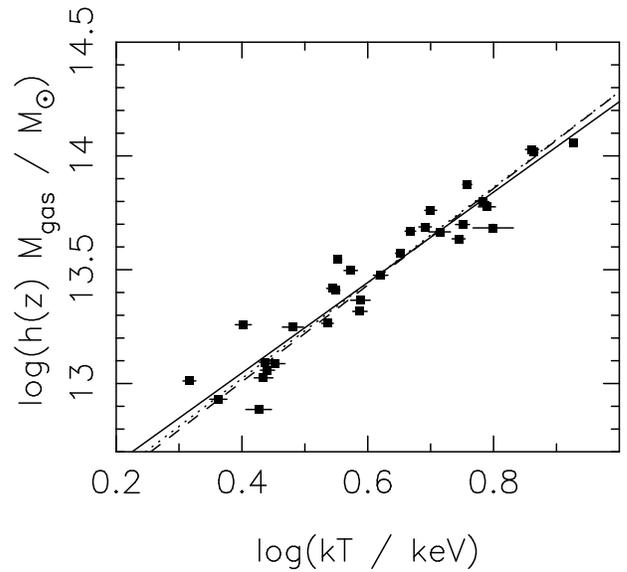}
\caption{Cluster gas mass vs. global temperature. Solid line is the
  best fitting relation to this sample obtained using WLSS, dashed
  line is the results obtained with BCES, and the dotted line is the
  best-fitting relation for the sample of regular clusters discussed
  in Arnaud et al. 2007.}
\label{gasmass}
\end{figure}

Fig.~\ref{gasmass} shows the relationship between gas mass and global
temperature for the REXCESS sample. Here we use temperature in the
$[0.15-0.75]R_{500}$ region to enable direct comparisons with previous
work. We fitted a scaling relation of the form $h(z)M_{gas} = C
[T_{X}/5 keV]^{\gamma}$ using two linear regression methods, a
weighted least squares method that incorporates an intrinsic scatter
term in the errors (WLSS), and BCES, which is an orthogonal method
that does not take into account the statistical errors. These methods
are discussed in more detail in Pratt et al. (2006). The results of
the fits are given in Table~\ref{mgast}. The slopes and normalisations
obtained with the two methods are consistent, and are also in good
agreement with those obtained for a smaller sample of relaxed clusters
by Arnaud et al. (2007). The results of our fits and the Arnaud et al.
relaxed cluster relation are shown in Fig~\ref{gasmass}. As shown in
Table~\ref{mgast}, the intrinsic logarithmic scatter for the REXCESS
sample is a factor of 2.5 times higher than for the Arnaud et al.
(2007) relaxed cluster sample. This is likely to arise as a result of
the wider range of cluster morphologies in our sample, although the
sample appears to include clusters with both higher and lower gas mass
for their temperature.

\begin{table*}
\centering
\caption{Results of linear regression analysis for the $M_{gas}$--$T$
relation. $\gamma$ is the slope in log-log space, $C$ the
normalisation, and $\sigma_{raw}$ and $\sigma_{intrins}$ the raw and
intrinsic logarithmic scatter about the relation, respectively.}
\label{mgast}
\begin{tabular}{llcccc}
\hline\hline
Sample & method&$\gamma$&$log(C)$&$\sigma_{raw}$&$\sigma_{intrins}$\\
\hline
REXCESS & WLSS&$1.986\pm0.111$&$13.652\pm0.020$&0.0928&0.0903\\
&BCES&$2.122\pm0.121$&$13.661\pm0.019$&0.0989&0.0962\\
Arnaud et al. (2007)&WLS&$2.10\pm0.05$&$13.65\pm0.01$&0.048&0.036\\
\hline
\hline
\end{tabular}
\end{table*}

\begin{figure*}
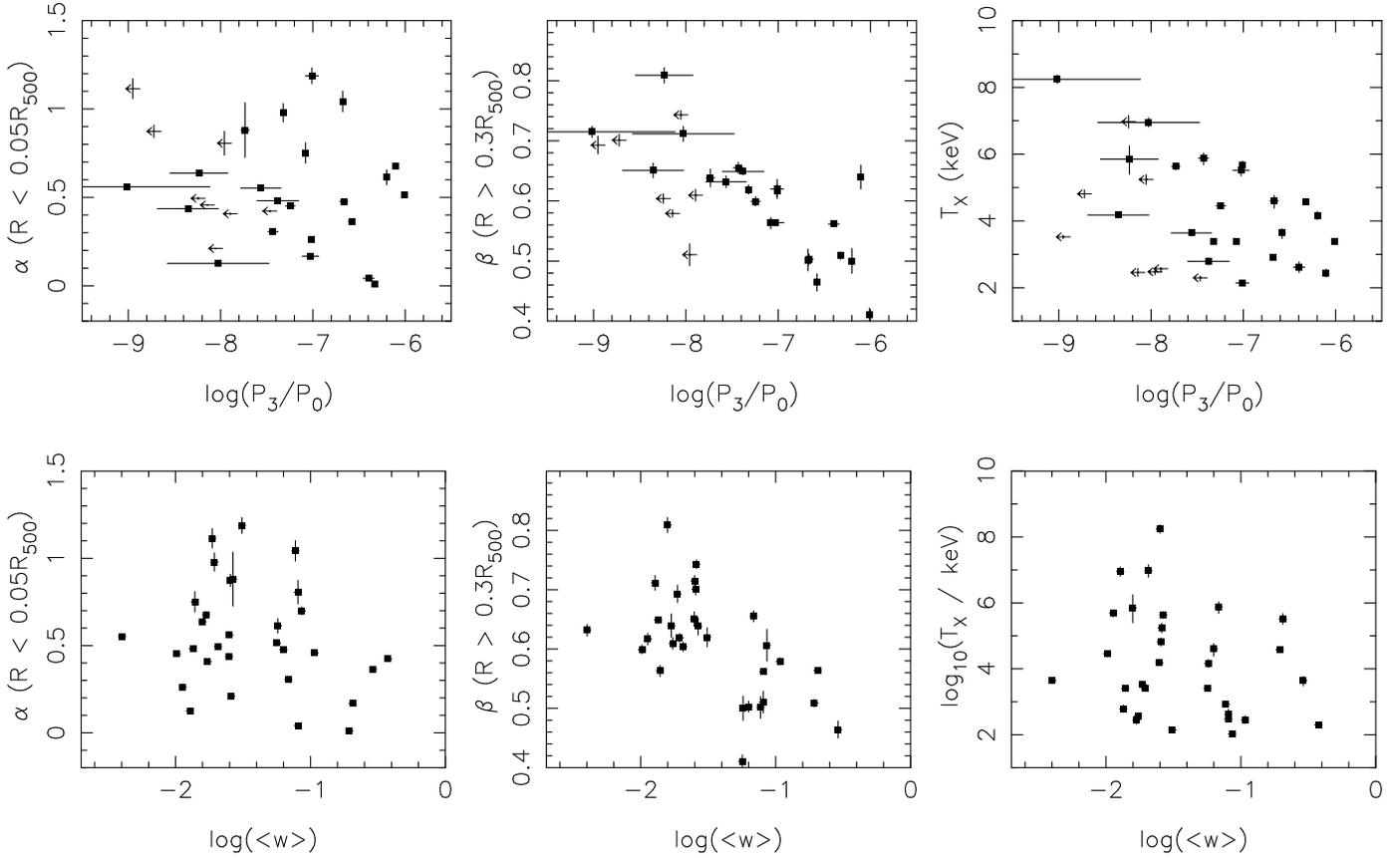

\centering
\resizebox{\hsize}{!} {
\epsfig{figure=9154f9a.eps}
\hspace{5mm}
\epsfig{figure=9154f9b.eps}
\hspace{5mm}
\epsfig{figure=9154f9c.eps}}

\vspace{5mm}
\resizebox{\hsize}{!} {
\epsfig{figure=9154f9d.eps}
\hspace{5mm}
\epsfig{figure=9154f9e.eps}
\hspace{5mm}
\epsfig{figure=9154f9f.eps}}
\caption{\label{pratios} The relationship between cluster dynamical
  state as parametrised by power ratios and centre shifts and gas
  properties. Top: $P_{3}/P_{0}$, bottom: $\langle w \rangle$. Left to
  right: inner slope of gas density ($\alpha_{<0.05}$), outer slope of
  gas density ($\beta_{0.3-0.8}$), temperature.}
\end{figure*}

We investigated the effect of choice of cluster centre on the gas mass
measurements and found that for the three most disturbed systems the
gas masses obtained varied by between 7 and 12\% for centroiding using
regions of radius 0.1, 0.3 and 0.5 $\times R_{500}$. As the cluster
dynamical state appears to be relatively independent of temperature
(see Section~\ref{dynstate}), we do not expect this systematic
uncertainty to affect our conclusions about the scaling of gas mass
and temperature. As, in addition, our sample selection strategy is
unbiased with respect to cluster dynamical state, the $\sim 9 \%$
scatter about the $M_{gas}$--$T$ relation we measure for this sample
should be a good measure of the intrinsic scatter in the relation for
the cluster population as a whole.

The steep relation between gas mass and temperature implies a decrease
in the total gas mass content of cooler clusters relative to higher
mass systems, when compared with standard predictions. This is clearly
connected with the excess of entropy seen at intermediate radii in
cooler systems and their relatively flat outer slopes. All of these
factors suggest that significant non-gravitational heating is likely
to have occurred, raising the gas entropy and lifting material beyond
$R_{500}$. Energetically it is most appealing if this energy injection
occurred prior to cluster collapse (eg., McCarthy et al 2007), but a
single level of ``preheating'' is unable to explain the full range of
observed properties, both as a function of cluster temperature and as
a function of the variations within in a temperture band. We will
explore the theoretical models that can account for the observed
trends in a future paper.

\subsection{Dependence of cluster structure on dynamical state}
\label{dynstate}

The radial gas distribution in galaxy clusters is likely to be
strongly affected by the cluster dynamical state, since mergers are
expected to disrupt the gas structure significantly. We used two
different quantitative measures of substructure in the 2-D surface
brightness distribution as means of characterising the cluster
dynamical state: the power ratio method of Buote \& Tsai (1995) and
centre shifts (e.g. Mohr et al. 1995, Poole et al. 2006).

The power ratio method used here is described in more detail in Pratt
et al. (2007). Here, we make use of power ratio measurements for the
entire REXCESS sample (to be discussed in a forthcoming paper),
determined in an aperture of radius $R_{500}$. We examined the
dependence of cluster structure on the three lowest order power ratios
of relevance for this work: $P_{2}/P_{0}$, which corresponds to a
measure of ellipticity, $P_{3}/P_{0}$, which is the best measure of
further substructure, and $P_{4}/P_{0}$, which also measures deviation
from a relaxed dynamical state. As some of the $P_{3}/P_{0}$
values are formally upper limits, we used the generalized Kendall's
$\tau$ test for censored data, as implemented in the {\sc asurv}
package. We did not find a correlation with inner gas density slope
($\alpha_{<0.05}$); however, there is evidence for a correlation
between $P_{3}/P_{0}$ and $\beta_{0.3-0.8}$, with a null hypothesis
probability of $\sim 0.1\%$ on the generalized Kendall's $\tau$ test.
There is weak evidence for a correlation between $P_{3}/P_{0}$ and
temperature, with a null hypothesis probability of $\sim 10\%$ on a
generalized Kendall's $\tau$ test. This correlation likely arises as a
result of the correlation between $\beta$ and temperature, as
discussion in Section~\ref{struc}. The relations between gas structure
and $P_{3}/P_{0}$ are shown in Fig.~\ref{pratios}.

\begin{figure*}
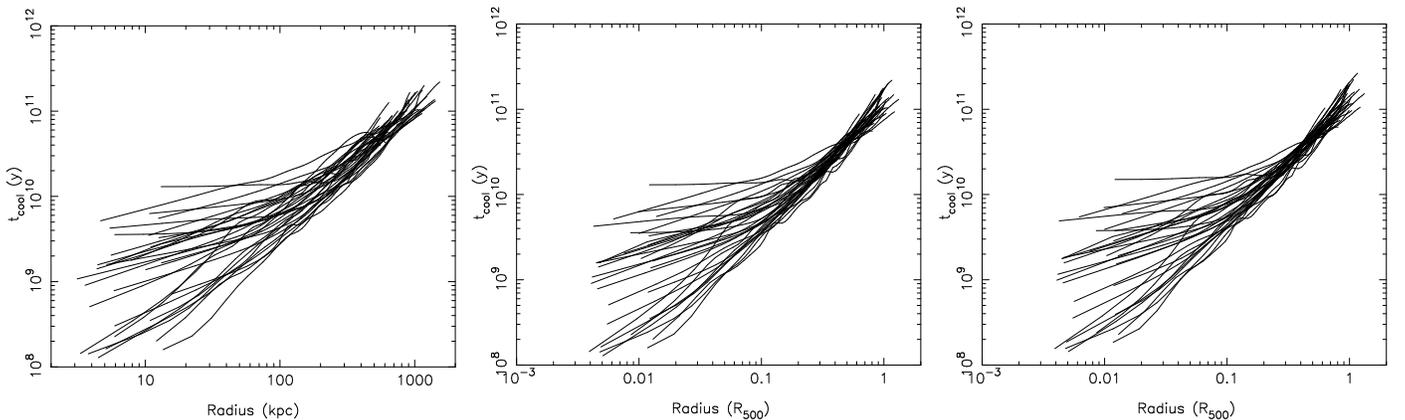

\centering 
\resizebox{\hsize}{!} {
\epsfig{figure=9154f10a.eps}
\hspace{4mm}
\epsfig{figure=9154f10b.eps}
\hspace{4mm}
\epsfig{figure=9154f10c.eps}}
\caption{\label{ctimes} Cooling time profiles for the cluster sample. Left: unscaled
  profiles in physical units; middle: unscaled gas density in units of
  scaled radius (corresponding to the top left panel of
  Fig.~\ref{nprofs}); right: profiles obtained from the
  redshift-scaled gas density profiles (corresponding to the middle
  left panel of Fig.~\ref{nprofs}).}
\end{figure*}

Centre shifts were determined for the entire REXCESS sample. Centroids
were obtained for regions of $n \times 0.1 \times R_{500}$ with
$n=1..10$ (i.e. the region with $r < 0.1 R_{500}$ was not included, so
that the centre shifts are not affected by the large scatter in
central gas properties). The results of this analysis will be
presented in more detail in a forthcoming paper; however, for the
purposes of a comparison with gas structural properties we used
$\langle w \rangle$, defined as the standard deviation of the
projected separations between the X-ray peak and centroid at each
radius (e.g. Maughan et al. 2007) in the region between $0.1 \times
R_{500}$ and $R_{500}$. We found that the exclusion of the central
region, possibly affected by a cooling core, did not significantly
affect the measured $\langle w \rangle$ values, except in the cases
where $\langle w \rangle$ is very small. In Fig.~\ref{pratios} we plot
$\langle w \rangle$ against inner and outer gas density slopes
($\alpha_{<0.05}$ and $\beta_{0.3-0.8}$). On a Spearman rank test, we
find no strong evidence for a significant correlation with
$\alpha_{<0.05}$, but find evidence for a correlation between $\langle
w \rangle$ and $\beta_{0.3-0.8}$, with a null hypothesis probabilty of
$0.08\%$ on a Spearman rank test. $\langle w \rangle$ is also
uncorrelated with temperature. The lack of correlation with inner
density slope suggests that cluster substructure and central cooling
behaviour are independent.

\begin{figure*}
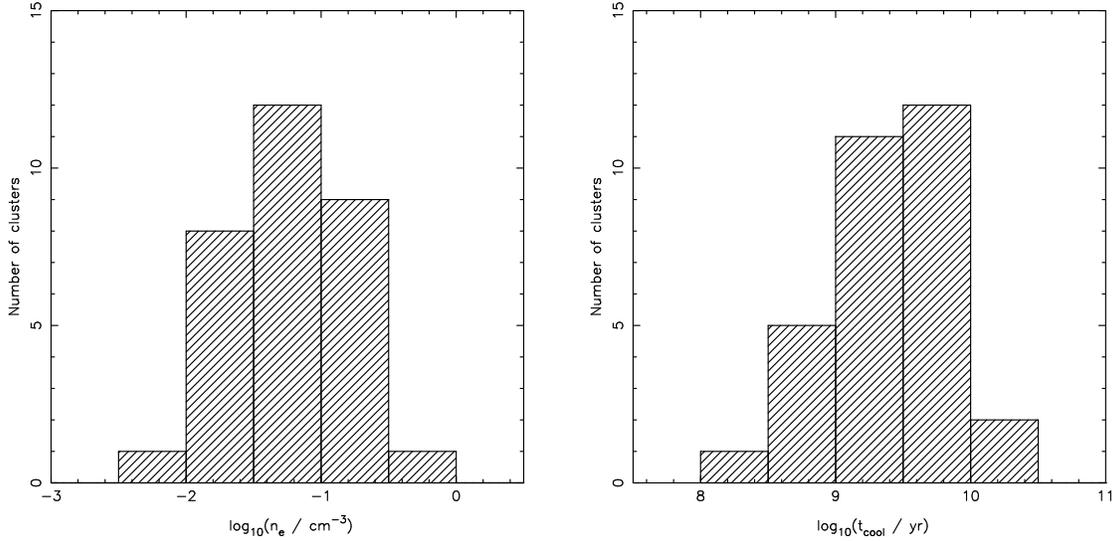

\centering
\resizebox{0.8\hsize}{!} {
\epsfig{figure=9154f11a.eps}
\hspace{2cm}
\epsfig{figure=9154f11b.eps}}
\caption{Histograms of central gas density and central cooling time
  (at $0.03R_{500}$, the innermost radius at which both density and
  temperature data is available for all clusters), showing that there
  is no strong evidence for bimodality in their cooling properties.}
\label{centreprop}
\end{figure*}

\subsection{Cooling times}

\begin{figure*}
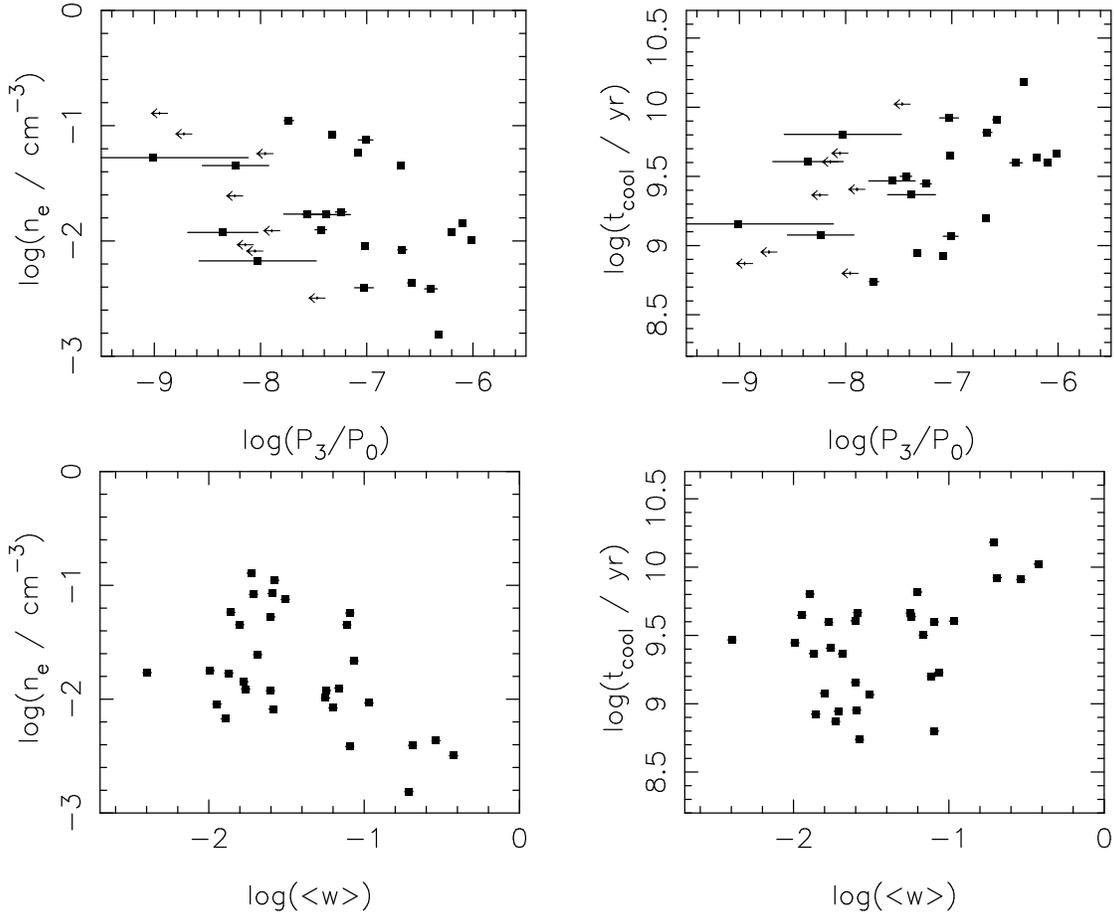

\centering
\resizebox{0.8\hsize}{!} {
\epsfig{figure=9154f12a.eps}
\hspace{2cm}
\epsfig{figure=9154f12b.eps}}
\resizebox{0.8\hsize}{!} {
\epsfig{figure=9154f12c.eps}
\hspace{2cm}
\epsfig{figure=9154f12d.eps}}
\caption{Comparisons between central gas properties and dynamical state. Left: $P_{3}/P_{0}$ vs. $n_{e} (0.007 R_{500}$) (top) and $t_{cool} (0.03 R_{500})$ (bottom); right: $\langle w \rangle$ vs. $n_{e} (0.007 R_{500}$) (top) and $t_{cool} (0.03 R_{500})$ (bottom).}
\label{cendyn}
\end{figure*}

Cooling time profiles were determined from the gas density and
temperature profiles for each cluster, where the cooling time,
$t_{cool}$ is defined as:
\begin{equation}
t_{cool} (r) = \frac{\frac{3}{2} \rho_{gas} (r) k T (r)
V}{L_{X} (r)}
\end{equation}
where $L_{X}(r)$ is the X-ray luminosity at radius $r$ determined
using the appropriate {\it mekal} model parameters at each radius in
{\sc xspec}. Fig.~\ref{ctimes} shows the cooling time profiles for all
31 clusters in physical and scaled units in the radial range where the
temperature profile is well constrained for each cluster. As the
temperature profiles we use here are not deprojected, we compared the
cooling time profiles obtained using projected and deprojected
temperature profiles for the cluster with the steepest central
temperature drop. We conclude that the use of projected temperatures
introduces an uncertainty of at most $\sim 15\%$ at radii less then
$0.02 R_{500}$ and has negligible effects at larger radii. 

\begin{figure}
\centering
\epsfig{figure=9154f13.eps,width=0.9\columnwidth}
\caption{Comparison of the observed gas density profiles (1$\sigma$
dispersion in red) and simulated profiles of Borgani et al. (2004)
(dashed lines) scaled by $\rho_{vir}$.}
\label{sim}
\end{figure}

\begin{figure*}
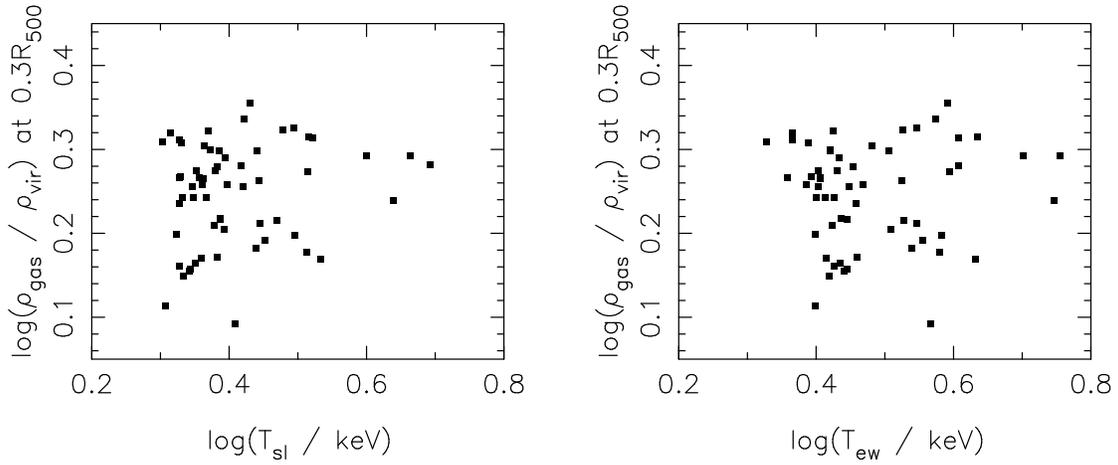

\centering
\resizebox{0.8\hsize}{!} {
\epsfig{figure=9154f14a.eps}
\hspace{2cm}
\epsfig{figure=9154f14b.eps}}
\caption{The gas density normalisation (at $0.3 R_{500}$) vs.
  temperature distribution for the simulated cluster sample using
  both the spectroscopic-like temperature (left) and the
  emission-weighted temperature (right). No significant correlation is
  seen}
\label{sim3}
\end{figure*}

\begin{figure*}
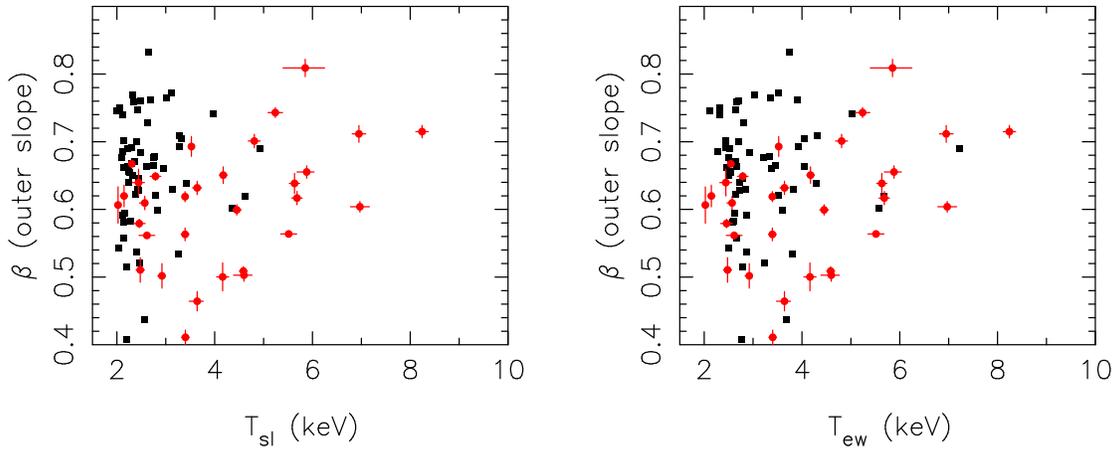

\centering
\resizebox{0.8\hsize}{!} {
\epsfig{figure=9154f15a.eps}
\hspace{2cm}
\epsfig{figure=9154f15b.eps}}
\caption{The outer slope ($\beta_{0.3-0.8}$) vs. temperature
  distribution for the observed and simulated cluster samples using
  both the spectroscopic-like temperature (left) and the
  emission-weighted temperature (right) for the simulated sample.}
\label{sim2}
\end{figure*}

The lower envelope of the cooling time profile distribution for
REXCESS is in good agreement with previous work (e.g. Sanderson et al.
2006, Bauer et al. 2005); however, the dispersion in the inner regions
of this sample is higher than seen in other samples, particularly
those dominated by cooling core clusters (e.g. Voigt \& Fabian 2004,
Sanderson et al. 2006), with a value of $\sigma/\langle t_{cool}
\rangle = 0.85$ at $0.03 R_{500}$ (as the cooling time profiles are
dominated by the behaviour of the gas density profiles, the scatter as
a function of radius is similar to that shown in Fig.~\ref{nprofs}).
We find a mean slope in the region $[0.2 - 0.8]R_{500}$ of
$1.4\pm0.3$, which is consistent with the results of Bauer et al.
(2005). In the region $[0.05-0.2]R_{500}$ the mean slope for our
sample is $\sim 0.9\pm0.3$, somewhat flatter than that seen over a
similar range in scaled radius by Sanderson et al. (2006), due to the
larger scatter towards higher cooling times in our sample. The spatial
resolution of our temperature profiles in the inner regions limits our
ability to draw strong conclusions about the central cooling
properties of this sample; however, the large dispersion in central
cooling properties suggests that there are more clusters with high
central cooling times in this sample compared to other studies.
Nevertheless, $t_{cool}$ drops below the Hubble time at a radius of
$\sim 50 - 200$ kpc for virtually all of the clusters in the sample.

Traditionally, clusters have been divided into two classes according
to their central cooling times: ``cool-core'' systems with short
central cooling times and high central densities and ``non-cool-core
systems'' with cooling times comparable to the Hubble time. Because of
their higher surface brightness, cool-core systems have tended to be
over represented in previous deep observations of clusters, leading to
the perception that the distribution of system cooling times is
bimodal. However, as shown in Fig.~\ref{centreprop}, the distributions
of gas densities and cooling times for the REXCESS sample at the
innermost radius appear to be well represented by a broad single
peaked distribution, and any division of this sample based on the
central cooling time would be arbitrary. We note, however, that our
cooling time profiles do not extend to very small radii, and so we
cannot rule out bimodality in the cooling time profiles on smaller
scales. In particular there does appear to be a subset of profiles
which continue to have a steep gradient in cooling time at small
radii, which corresponds primarily to those profiles with $t_{cool}
(0.03 R_{500}) < 1$ Gyr. The mean slope for profiles with inner
cooling times higher than 1 Gyr is significantly flatter, with several
profiles that are asymptotically flat in the central regions. Hence
our data do not rule out models in which conduction can stabilize the
cores of clusters with high central cooling times (e.g. Donahue et al.
2005). As noted in Section~\ref{struc}, the distribution of inner gas
density slope ($\alpha_{<0.05}$) also shows no strong evidence
for bimodality. These conclusions are independent of the choice of
radius and gas density scaling.

Finally, we investigated the connection between central gas properties
and cluster dynamical state by comparing the inner cooling times and
gas densities with $\langle w \rangle$ and $P_{3}/P_{0}$. Although it
is widely anticipated that clusters with long central cooling times
result from cluster mergers, recent papers have suggested that central
density cusps are unlikely to be destroyed in the merger event (eg.,
McCarthy et al.\ 2004, Borgani et al.\ 2004, Poole et al., 2006). It
is therefore of considerable interest to investigate the connection
between central gas properties and cluster dynamical state by
comparing the inner cooling times and gas densities with $\langle w
\rangle$ and $P_{3}/P_{0}$. Fig.~\ref{cendyn} illustrates that
correlations are present, in the sense that clusters with lower
central densities, and higher central cooling times, tend to show more
evidence for disturbance. However, the trend is driven by a few
systems with the highest level of disturbance, and the null hypothesis
that there is no correlation can only be rejected at the $\sim 85 - 97
\%$ level using the Spearman rank correlation test (for $\langle w
\rangle$) or the generalized Kendall's $\tau$ test (for
$P_{3}/P_{0}$). While there is therefore weak evidence that merger
activity can affect cluster central densities and cooling times, it
does not appear to affect the slope of the density profile
(Section~\ref{dynstate}) in the inner regions.  Clearly these issues
need to be addressed by careful comparison with cosmological
simulations.

\subsection{Comparison with simulations}

We compared the observed gas density profiles with the simulated
profiles of all clusters with $kT > 2$ keV from Borgani et al. (2004),
in which the SPH code GADGET-2 (Springel 2005) was used to simulate a
concordance $\Lambda$CDM model ($\Omega_M = 0.3$, $\Omega_{\Lambda} =
0.7$, $\sigma_{8} = 0.8$, $h = 0.7$) within a box of 192 $h^{-1}$ Mpc
on a side, using $480^{3}$ dark matter particles and an equal number
of gas particles. The simulation included radiative cooling, star
formation and galactic ejecta powered by supernova feedback.
Fig.~\ref{sim} shows in red the $1\sigma$ dispersion of the entire
sample of observed profiles as in the bottom left panel of
Fig.~\ref{nprofs}, with the simulated profiles as dashed lines.
Observed profiles were scaled by $\rho_{vir}$, defined as $100
\rho_{crit,0}$. The agreement is very good in the radial range
$[0.02-0.3]R_{500}$; however at larger radii the observed profiles are
slightly flatter than the simulated profiles. We found that the
simulated profiles have a mean slope in the radial range
$[0.3-0.8]R_{500}$ of $-1.97\pm0.25$, which is consistent within the
errors with our measured mean slope of $-1.80\pm0.28$, but is slightly
higher. For the purpose of this comparison we used the ``true''
$R_{500}$ values from the simulations. The small difference between
the slope of the observed and simulated profiles at large radius may
be due to differences in the definition of $R_{500}$. We cannot apply
our method for determining $R_{500}$ based on the observed
$Y_{X}-M_{500}$ relation directly to the simulations, because
spectroscopic-like temperature measurements that exclude the central
regions of the cluster are not available. The relative dispersion in
the simulated profiles at $0.3R_{500}$ is $\sigma/\langle n_{e}
\rangle = 0.13$, which is significantly lower than the value of $\sim
0.22$ for the redshift-scaled observed profiles. 

Nagai et al. (2007) carried out a similar comparison of the gas
  distributions in simulated and observed clusters, using an observed
  sample consisting of the $z=0$ relaxed cluster sample of Vikhlinin
  et al. (2006). They found slightly better agreement in the outer
  regions between the {\it Chandra} profiles and their simulations
  that included cooling and star formation. Their slightly better
  agreement between observations and simulations may be a result of
  comparing with a sample including only relaxed clusters. Differences
  in the numerical treatments and implementations of cooling and star
  formation in the simulations may also be relevant.

We also investigated whether the temperature-dependence of gas density
normalisation seen in the REXCESS clusters (see Section~\ref{tdep}) is
also present in the simulations. Figure~\ref{sim3} shows the
relationship between gas density normalisation and temperature (shown
for both the emission-weighted and spectroscopic-like temperatures)
for the simulations. There is no significant temperature-dependence of
gas density normalisation for the simulated cluster sample.

Finally we compared the outer slope with gas temperature, using both
the emission-weighted temperatures within $R_{500}$ and the
spectroscopic-like temperatures defined in Rasia et al. (2005).
Fig.~\ref{sim2} shows that the correlation between gas density slope
($\beta_{0.3-0.8}$) and temperature observed in the REXCESS sample is
not present in the simulated data (on a Spearman rank test we find
null hypothesis probabilities of $35\%$ and $75\%$ for $T_{sl}$ and
$T_{ew}$, respectively, compared with $<1\%$ for the observed sample)
-- there are clearly a number of simulated clusters in the low $T$,
steep $\beta$ region of Fig.~\ref{sim2}, which is not populated by the
observed sample. However, the temperature measurements for the
simulated sample do not exclude the central region which makes it
difficult to draw firm conclusions. It is also clear from
Fig.~\ref{sim2} that the temperature distribution in the simulated
sample is different to that of the observed sample, with fewer high
temperature clusters.

These differences hint that the simulations may not match the true
thermal history of the intracluster medium. In particular, the heat
input appears too centrally concentrated so that the excess entropy
seen in the lower temperature clusters at intermediate radii is not
reproduced. A more detailed comparison with simulations will form part
of a later paper. 

\section{Conclusions}

We have presented the first detailed study of the structural
properties of cluster gas in a large, representative sample of nearby
galaxy clusters. As the sample was selected by X-ray luminosity, it
includes clusters of all dynamical states, allowing us to investigate
the effect that the inclusion of less regular systems has on results
obtained with previous studies of regular clusters. We found the
following results:
\begin{itemize} 
\item The 1-D gas density profiles scale self-similarly, with a
  scatter ranging from $\sim 100\%$ in the inner regions to $\sim 20
  \%$ at a radius of $0.3R_{500}$ when expected evolution with
  redshift is taken into account.
\item Gas density normalisation at $0.3 R_{500}$ is correlated with
  global cluster temperature, with a scaling of $n_{e} (0.3R_{500})
  \propto T_{X}^{0.5}$, consistent with the expectation of modified
  entropy-temperature scaling models. Using this scaling reduced the
  scatter in the gas density profiles at $0.3 R_{500}$ to $\sim 15
  \%$; however, the scatter at larger radii is slightly increased,
  which indicates that the entropy excess is much less significant
  beyond $\sim 0.5 R_{500}$.
\item The gas density slope continues to increase with
  radius in the region $0.5R_{500} - R_{500}$, as found by others,
  which is of importance for cluster mass estimates.
\item The outer gas density slope is correlated with X-ray
  temperature, primarily due to a lack of hot clusters with flat gas
  distributions. The flatter slope in lower temperature systems,
  combined with the entropy excess at intermediate radii and the steep
  slope of the $M_{gas}$--$T$ relation suggests that gas has been
  displaced from the centres of the lower temperature systems to
  larger radii.
\item Based on a characterisation of the cluster dynamical state using
  power ratios and centre shifts, there is evidence of a
  correlation between cluster dynamical state and outer gas
  density slope, and no correlation with inner gas density slope.
  There is also evidence for a correlation between dynamical
  state and the central gas properties (gas density normalisation at
  $0.007 R_{500}$ and cooling time at $0.03 R_{500}$).
\item There is no evidence for bimodality in the central gas density,
  gas density slope or cooling times for this sample, suggesting that
  X-ray clusters form a single population with a continuous
  distribution of central gas properties.
\item The gas-mass temperature relation for the REXCESS sample is in
good agreement with predictions of self-similar scaling modified by
the presence of an entropy excess, and with previous work on samples
of regular clusters; however, the intrinsic scatter is a factor of
$\sim 2.5$ times higher than for the relaxed cluster population. 
 \item The scaling properties of the gas density profiles appear to
  be in broad agreement with those of simulated cluster samples
  (Borgani et al. 2004) at intermediate radius, with a slightly
  flatter slope in the outer regions. However, in contrast to the
  observational data there is no correlation between gas density
  normalisation and temperature in the simulated sample, or between
  the outer slope and temperature. These discrepancies suggest that
  the non-gravitational heating of the intra-cluster medium may be too
  centrally concentrated in these models.

\end{itemize}

\begin{acknowledgements}
We are grateful for helpful comments and discussion with members of
the REXCESS collaboration, particularly Stefano Borgani, Chris
Collins, Alexis Finoguenov, Thomas Reiprich, and Kathy Romer. We thank
the referee for helpful suggestions.
\end{acknowledgements}

\section*{Appendix 1: Surface brightness and density profiles}
\begin{figure*}
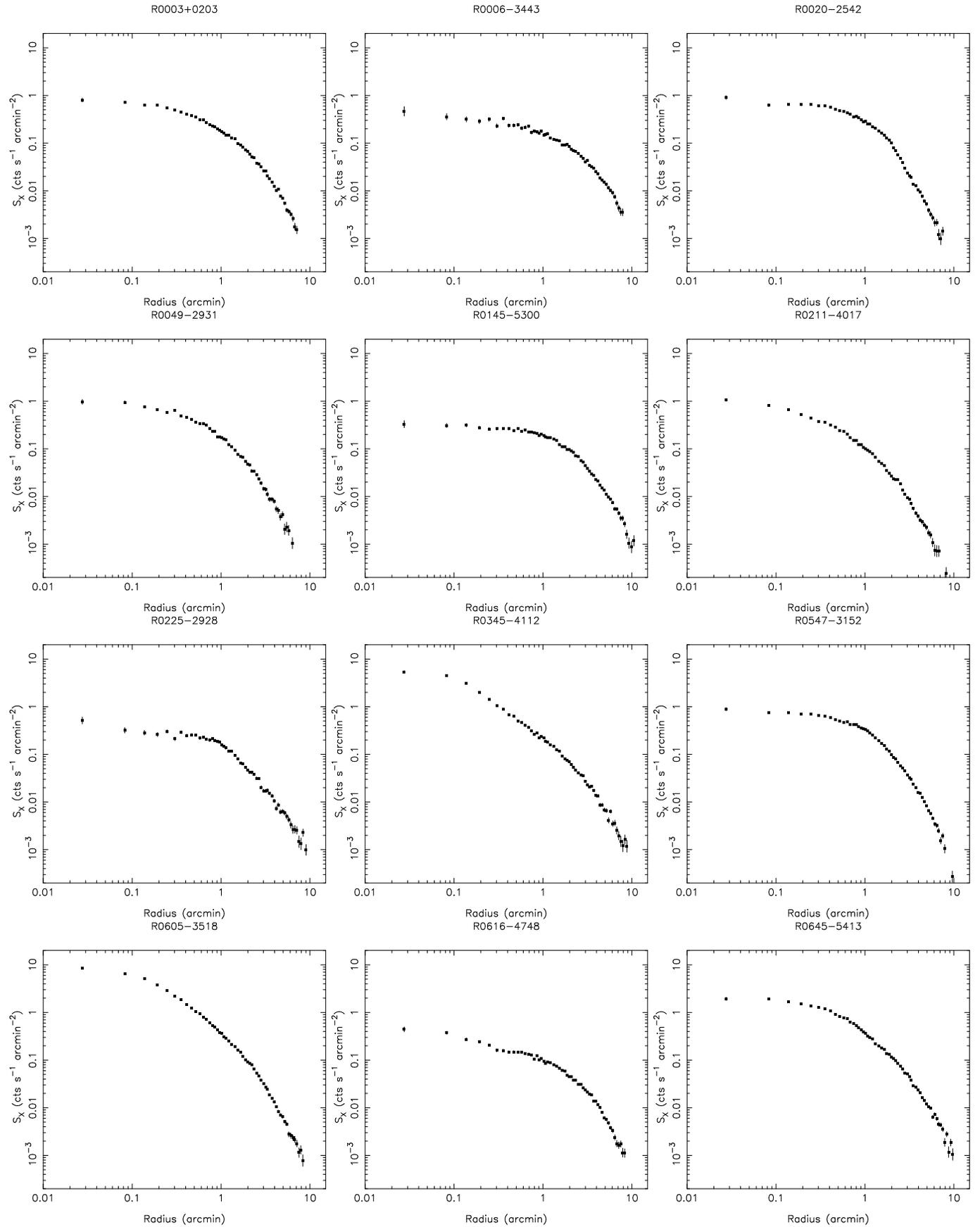

\centering
\vbox{
\hbox{
\epsfig{figure=9154f16a.eps,width=6.0cm}
\epsfig{figure=9154f16b.eps,width=6.0cm}
\epsfig{figure=9154f16c.eps,width=6.0cm}
}
\hbox{
\epsfig{figure=9154f16d.eps,width=6.0cm}
\epsfig{figure=9154f16e.eps,width=6.0cm}
\epsfig{figure=9154f16f.eps,width=6.0cm}
}
\hbox{
\epsfig{figure=9154f16g.eps,width=6.0cm}
\epsfig{figure=9154f16h.eps,width=6.0cm}
\epsfig{figure=9154f16i.eps,width=6.0cm}
}
\hbox{
\epsfig{figure=9154f16j.eps,width=6.0cm}
\epsfig{figure=9154f16k.eps,width=6.0cm}
\epsfig{figure=9154f16l.eps,width=6.0cm}
}
}
\caption{Co-added MOS1, MOS2 and pn surface brightness profiles for
  the entire sample in the energy band 0.3 - 2.0 keV.}
\label{sxprofs1}
\end{figure*}

\begin{figure*}[!b]
\centering
\vbox{
\hbox{
\epsfig{figure=9154f17a.eps,width=6.0cm}
\epsfig{figure=9154f17b.eps,width=6.0cm}
\epsfig{figure=9154f17c.eps,width=6.0cm}
}
\hbox{
\epsfig{figure=9154f17d.eps,width=6.0cm}
\epsfig{figure=9154f17e.eps,width=6.0cm}
\epsfig{figure=9154f17f.eps,width=6.0cm}
}
\hbox{
\epsfig{figure=9154f17g.eps,width=6.0cm}
\epsfig{figure=9154f17h.eps,width=6.0cm}
\epsfig{figure=9154f17i.eps,width=6.0cm}
}
\hbox{
\epsfig{figure=9154f17j.eps,width=6.0cm}
\epsfig{figure=9154f17k.eps,width=6.0cm}
\epsfig{figure=9154f17l.eps,width=6.0cm}
}
}
\caption{Co-added MOS1, MOS2 and pn surface brightness profiles for
the entire sample (cont.)}
\label{sxprofs2}
\end{figure*}

\begin{figure*}[!b]
\centering
\vbox{
\hbox{
\epsfig{figure=9154f18a.eps,width=6.0cm}
\epsfig{figure=9154f18b.eps,width=6.0cm}
\epsfig{figure=9154f18c.eps,width=6.0cm}
}
\hbox{
\epsfig{figure=9154f18d.eps,width=6.0cm}
\epsfig{figure=9154f18e.eps,width=6.0cm}
\epsfig{figure=9154f18f.eps,width=6.0cm}
}
\hbox{
\epsfig{figure=9154f18g.eps,width=6.0cm}
}
}
\caption{Co-added MOS1, MOS2 and pn surface brightness profiles for
the entire sample (cont.)}
\label{sxprofs3}
\end{figure*}
\begin{figure*}[!b]
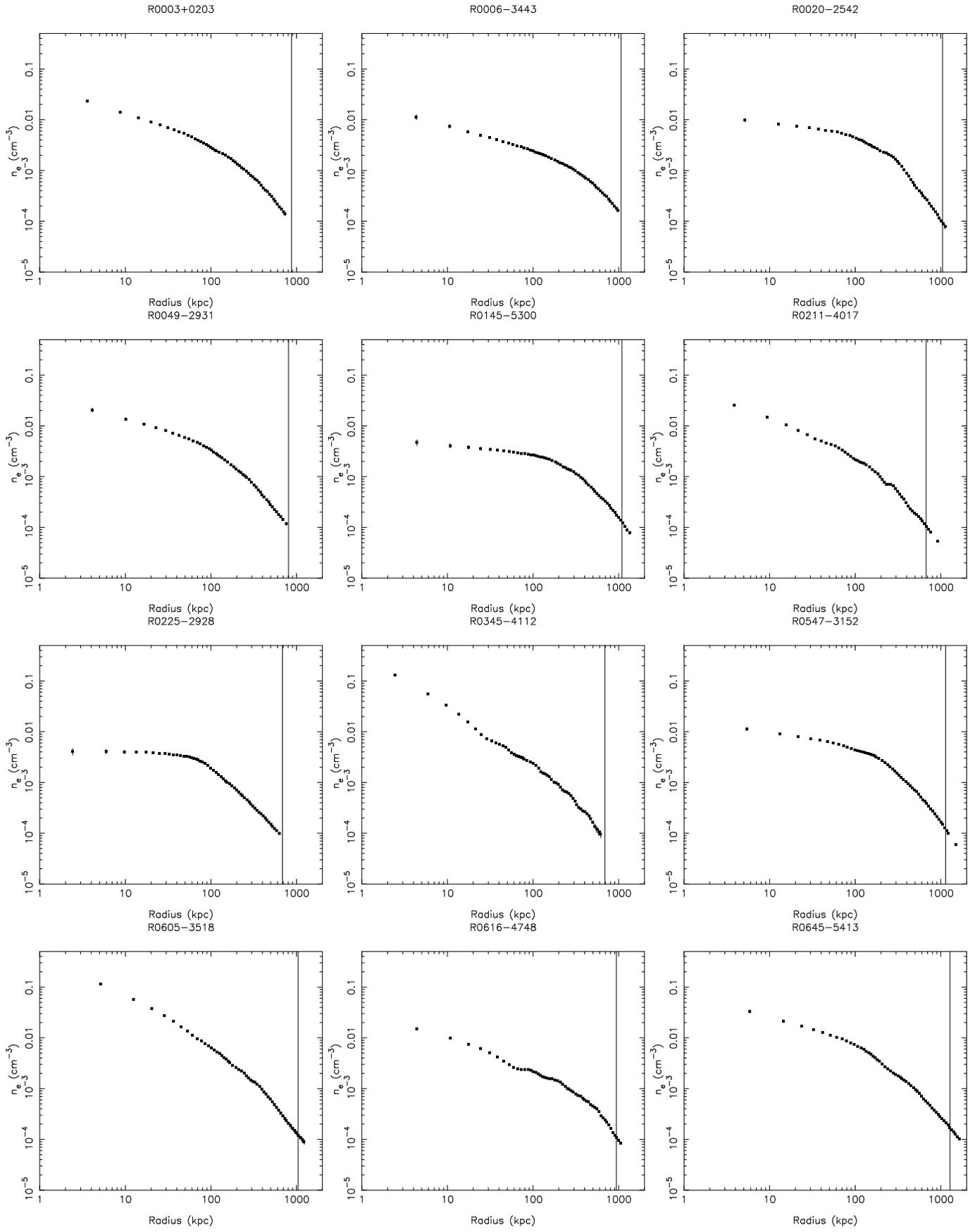

\centering
\vbox{
\hbox{
\epsfig{figure=9154f19a.eps,width=6.0cm}
\epsfig{figure=9154f19b.eps,width=6.0cm}
\epsfig{figure=9154f19c.eps,width=6.0cm}
}
\hbox{
\epsfig{figure=9154f19d.eps,width=6.0cm}
\epsfig{figure=9154f19e.eps,width=6.0cm}
\epsfig{figure=9154f19f.eps,width=6.0cm}
}
\hbox{
\epsfig{figure=9154f19g.eps,width=6.0cm}
\epsfig{figure=9154f19h.eps,width=6.0cm}
\epsfig{figure=9154f19i.eps,width=6.0cm}
}
\hbox{
\epsfig{figure=9154f19j.eps,width=6.0cm}
\epsfig{figure=9154f19k.eps,width=6.0cm}
\epsfig{figure=9154f19l.eps,width=6.0cm}
}
}
\caption{Individual gas density profiles for the entire sample.
  Vertical lines indicate $R_{500}$ for each cluster.}
\label{neprofs1}
\end{figure*}

\begin{figure*}[!b]
\centering
\vbox{
\hbox{
\epsfig{figure=9154f20a.eps,width=6.0cm}
\epsfig{figure=9154f20b.eps,width=6.0cm}
\epsfig{figure=9154f20c.eps,width=6.0cm}
}
\hbox{
\epsfig{figure=9154f20d.eps,width=6.0cm}
\epsfig{figure=9154f20e.eps,width=6.0cm}
\epsfig{figure=9154f20f.eps,width=6.0cm}
}
\hbox{
\epsfig{figure=9154f20g.eps,width=6.0cm}
\epsfig{figure=9154f20h.eps,width=6.0cm}
\epsfig{figure=9154f20i.eps,width=6.0cm}
}
\hbox{
\epsfig{figure=9154f20j.eps,width=6.0cm}
\epsfig{figure=9154f20k.eps,width=6.0cm}
\epsfig{figure=9154f20l.eps,width=6.0cm}
}
}
\caption{Individual gas density profiles for the entire sample (cont.)}
\label{neprofs2}
\end{figure*}

\begin{figure*}[!b]
\centering
\vbox{
\hbox{
\epsfig{figure=9154f21a.eps,width=6.0cm}
\epsfig{figure=9154f21b.eps,width=6.0cm}
\epsfig{figure=9154f21c.eps,width=6.0cm}
}
\hbox{
\epsfig{figure=9154f21d.eps,width=6.0cm}
\epsfig{figure=9154f21e.eps,width=6.0cm}
\epsfig{figure=9154f21f.eps,width=6.0cm}
}
\hbox{
\epsfig{figure=9154f21g.eps,width=6.0cm}
}
}
\caption{Individual gas density profiles for the entire sample (cont.)}
\label{neprofs3}
\end{figure*}

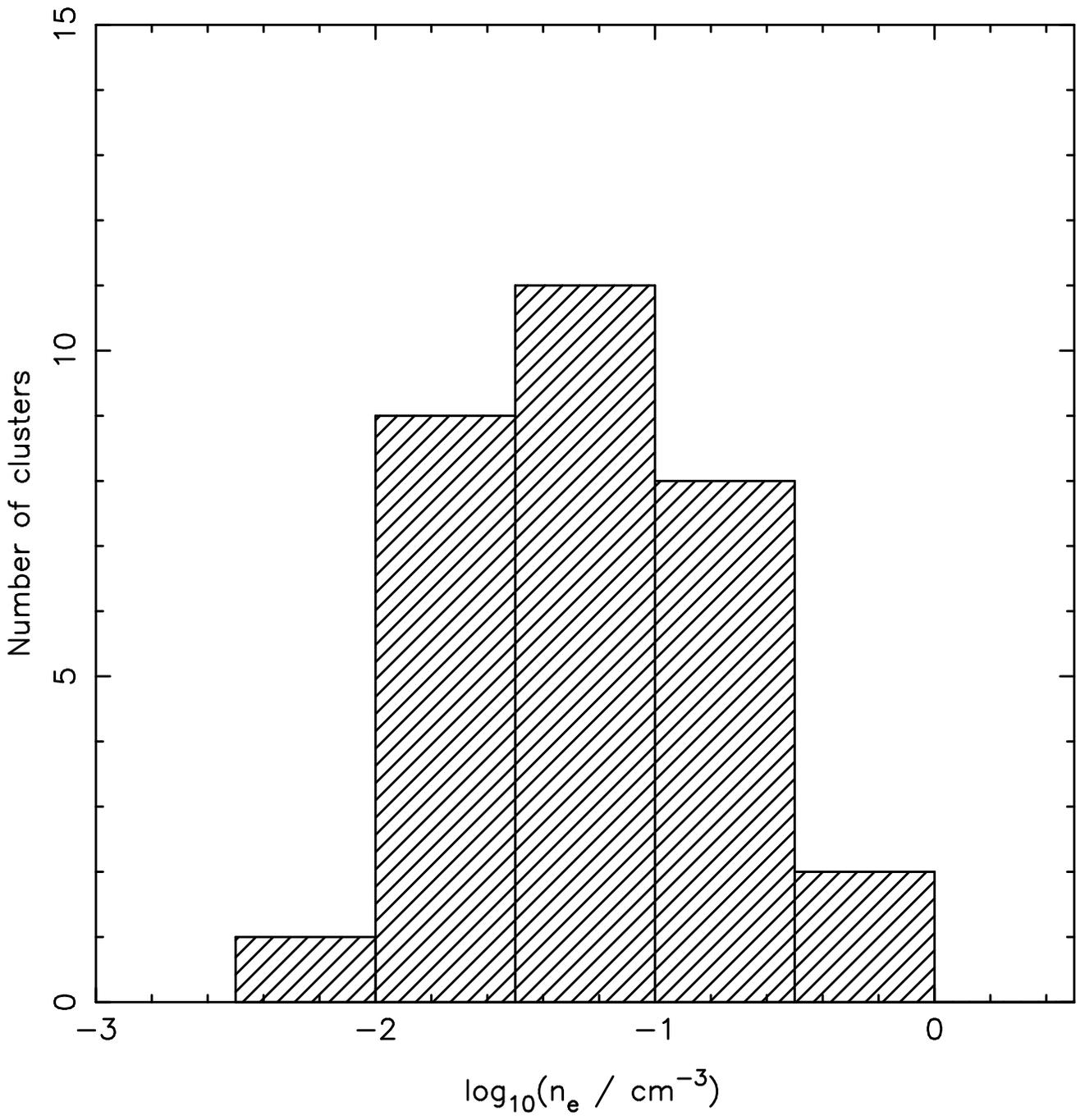
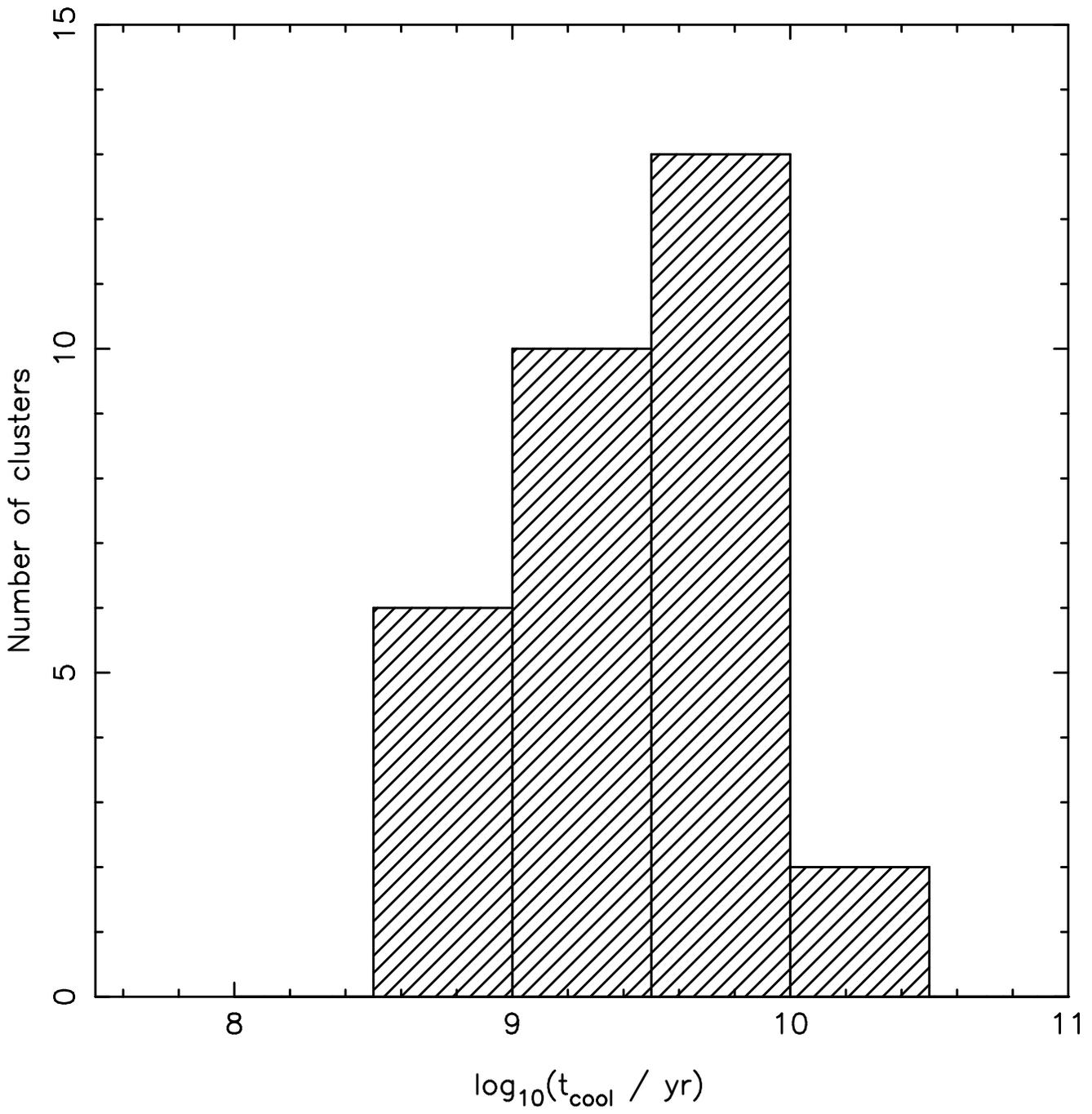
\end{document}